\documentclass{article}
\usepackage{epsfig}

\date{\today}

\begin{document}
\begin{center}
{\Large\bf BOSON STARS WITH NEGATIVE COSMOLOGICAL CONSTANT}
\\[0.8cm]
Dumitru Astefanesei\footnote{Present address: Perimeter Institute for Theoretical Physics, Waterloo, Ontario N2J 2W9, Canada}
\\
{\small \emph{Department of Physics, McGill University, Montreal, QC, H3A 2T8, Canada}
\\
e-mail: astefand@physics.mcgill.ca}
\\[0.2cm]
and 
\\[0.2cm]
Eugen Radu
\\
{\small \emph{Albert-Ludwigs-Universit\"at Freiburg, Fakult\"at f\"ur Physik}, 
\\
\emph{Hermann-Herder-Stra\ss e 3, D-79104 Freiburg, Germany}
\\ email: radu@newton.physik.uni-freiburg.de}
\\[0.8cm]
\end{center}
\begin{abstract}
We consider boson star solutions in a $D$-dimensional, 
asymptotically anti-de Sitter spacetime
and investigate the influence of the cosmological term
on their properties.
We find that for $D>4$ the boson star properties are 
close to those in four dimensions with a vanishing cosmological constant.
A different behavior is noticed for the solutions in the three dimensional case.
We establish also the non-existence of static, spherically symmetric black holes 
with a harmonically time-dependent complex scalar field in any dimension greater than two.
\end{abstract}

\section{INTRODUCTION}
 
Recently a tremendous amount of interest has been focused on anti-de Sitter (AdS) spacetime.
This interest is mainly motivated by the proposed correspondence between physical effects 
associated with gravitating fields propagating in AdS spacetime and those of a conformal
field theory on the boundary of AdS spacetime \cite{Witten:1998qj,Maldacena:1997re}.

Being a maximally symmetric spacetime, AdS spacetime is also an excellent 
model to investigate questions of principle related to the quantisation of fields
propagating on a curved background, the interaction with the gravitational field
and issues related to the lack of global hyperbolicity.

In view of these developments, an examination of the classical solutions of gravitating 
fields in asymptotically AdS (AAdS) space seems appropriate.

Recently, some authors have discussed the properties 
of gravitating SU(2) nonabelian fields for $\Lambda <0$ 
\cite{Winstanley:1998sn,Bjoraker:2000qd,Radu:2002ij}.
They obtained some surprising results, which are strikingly different from 
the results found in the asymptotically flat case
(for example the existence of regular dyon solutions without a Higgs field).
Insofar as the no-hair theorem is concerned, it has been shown that 
there exist stable black hole solutions in 
SU(2) Einstein-Yang-Mills (EYM) theory that are AAdS \cite{Winstanley:1998sn,Bjoraker:2000qd}.
However, the same system in the presence of a Higgs scalar
presents solutions with very similar properties to the asymptotically-flat space counterparts
\cite{Lugo:1999fm, Lugo:2000ai, VanderBij:2001ah}.
More recently it has been shown that the $U(1)$ Higgs field equations have a 
vortex solution in four-dimensional AAdS \cite{Dehghani:2001ft}.
Static spherically symmetric solutions of general relativity with $\Lambda<0$,
coupled to a real scalar field, have been studied in \cite{Das:2001rk}.
Surprisingly enough, the solutions have finite energy although 
there are naked singularities 
and present interesting properties in the context of the AdS/CFT correspondence.

All of these studies assume that the matter fields 
have the same symmetries as the underlying spacetime;
in particular, the scalar fields are assumed to be invariant under the action
of the stationary Killing vector.
However, the consideration of static configurations in which a complex scalar 
field varies harmonically with time, 
with a static energy-momentum tensor, yields new kinds
of regular configurations, the so-called 'boson star' solutions.

The study of boson stars can be traced back to the work of Kaup \cite{Kaup} and 
Ruffini and Bonazzalo  \cite{Ruffini:1969qy} more than 30 years ago.
They found asymptotically flat, spherically symmetric equilibrium solutions
of the Einstein-Klein-Gordon equations.
These boson stars are macroscopic quantum states and are only prevented 
from collapsing
gravitationally by the Heisenberg uncertainty principle.
Although boson stars have many similarities to their fermionic counterparts,
there are many interesting differences.
For example, boson stars also exhibit a critical mass and critical particle number.
Later works considered the more complicated possibilities of the presence 
of a self interaction
for the scalar field \cite{Mielke:1980sa,Colpi:1986ye,Schunck:1999zu}
or a nonminimal coupling of the scalar field to gravity \cite{vanderBij:1987gi}.
Jetzer and van der Bij extended the model to include a $U(1)$ gauge charge \cite{Jetzer:1989av}
(see also \cite{Dariescu:cx}).
Boson stars in the presence of a dilaton or an axidilaton have 
also been studied by various authors \cite{Gradwohl:1989px}, 
as well as boson-fermion stars \cite{Henriques:1989ez}. 

All these models have demonstrated the same characteristic: new interactions tends to 
increase the critical values
of mass and particle number, although the particular values are very model dependent.
The stability against perturbations 
around the equilibrium state has been discussed also  by a number of authors 
\cite{Gleiser:1988rq}-\cite{Kusmartsev:cr}.
An extensive review of the boson star properties is given in \cite{Jetzer:1992jr}, \cite{Mielke:1997re}.

All of these solutions assumed a spherically symmetric ansatz; a more recent development 
is the study 
of axially symmetric boson star configurations \cite{Schupp:1995dy,Yoshida:1997jq,Yoshida:nd}.
Asymptotically flat, rotating boson stars have been obtained in \cite{s1,Yoshida:1997qf}.

The recent interest in these solitonic objects is largely due to
the suggestion that the dark matter could be bosonic in nature.
With scalar fields often used to model early universe, there is the possibility
that such solutions could condense to form boson stars.

Most of the works on boson star properties have been carried out on the 
assumption that spacetime is asymptotically flat.
Because of the physical importance of these objects, it is worthwhile to study
generalizations in a different cosmological background.
We may hope that similar to a nonabelian theory, the different asymptotic structure 
of the spacetime will affect the properties of the solutions leading to some 
new effects. 

The goal of this paper is to reexamine the basic properties of soliton stars 
in the presence of a negative cosmological constant. 
The solutions we are looking for are the asymptotically 
AdS analogues of the Ruffini and Bonazzalo configurations \cite{Ruffini:1969qy}.
To our knowledge, a discussion of this case is not available in the literature.

We find that the basic properties of a four dimensional soliton star will remain basically 
the same even in the presence of a negative cosmological constant.
However a nonzero $\Lambda$  term in the action implies a complicated power decay of 
the fields at infinity, rather than exponentially.
Also, as expected, the parameter of the solutions found in Ref.\cite{Ruffini:1969qy}
(mass, particle number, effective radius)
remains no 
longer valid and new values are found for every value of $\Lambda$. 
The main effect of a nonzero cosmological constant is to decrease the 
maximal mass of the star.

Motivated by recent interest in higher dimensional gravity with negative cosmological constant,
we look
also for spherically symmetric boson stars with $\Lambda \leq 0$ 
in spacetimes with a number of dimensions $D \neq 4$.  
It is always of interest to see how the dimensionality of spacetime affects the physical 
consequences of a given theory.

The higher-dimensional boson star solutions  might be relevant in the context of superstring theory
as well as for understanding how the behaviour depends on the dimensionality
of the spacetime.
In particular we are interested in the five-dimensional case.
To our knowledge very little is known about the properties of these type of configurations
in $D \neq 4$, although the exact boson star solutions in three-dimensional gravity with $\Lambda<0$
were found in \cite{Sakamoto:1998hq}.
A qualitative discussion  of the boson star properties in the context of brane world 
models with compact large extra dimensions has been done in \cite{Stojkovic:2001qi}.
As argued there, the properties of large soliton stars will
remain similar even in the context of the brane world and large extra dimensions scenarios.
However, at distances smaller than the compactification radius,
the properties of a boson stars are quite different than in the $(3+1)$-dimensional case.

We find that for $D>3$ and a negative cosmological constant, the properties of the solutions
are similar to the well-known four-dimensional asymptotically flat case.
The total mass parameter $M$ rises quickly with the initial value of the scalar field at the origin
$\varphi_0$ to a maximum,
then drops and has several local minima and maxima until it approaches an asymptotic
value for large $\varphi_0$. A similar behaviour is found for the particle number $N$.
The solutions for a vanishing cosmological constant are discussed as a particular
case, since they present the same general pattern.
A separate section is dedicated to the rather special case $D=3$. 
Exact solutions in this case (in the limit of large self interaction) are know to exist.
However, the properties of the boson star solutions in $AdS_3$ spacetime   
are different as compared to the higher dimensional case.
We notice the existence of a maximal allowed value for the central density and 
the absence of local extrema for $N$ and $M$. 
Also, no solutions are found in the asymptotically flat limit.

One may ask about the possible relevance of these macroscopic quantum states 
within the AdS/CFT correspondence.
Since the proposal of Maldacena's conjecture, which gives a correspondence between a
theory of gravity in $AdS_D$ spacetime and a field theory on its $(D-1)$-dimensional boundary, much intense
work has been devoted to get a deeper understanding of its implications.
The AdS/CFT correspondence for a scalar field has been discussed by a number of authors, 
however without considering
these type of time-dependent solutions. 
We present in this paper only a preliminary approach to this question. 
The boundary stress tensor
and the associated conserved charges are computed for three, four and five dimensions.
In three and five dimensions, the counterterm prescription \cite{Balasubramanian:1999re}
gives an additional vacuum 
(Casimir) energy, which agrees with that found in the context of AdS/CFT correspondence.

The paper is structured as follows:  in Section 2 we present the general   
framework and analyse the field equations and boundary conditions exactly.
In Section 3 we address the question of AAdS black hole solutions with complex scalar hair.
In $4D$ asymptotically flat space, the boson stars 
do not allow for black hole horizon inside them \cite{Pena:1997cy}.
We generalize this result to include a negative $\Lambda$, for any $D\geq 3$.
In Section 4 the field equations are solved numerically.
It is shown how the dimensionality of spacetime affects the properties of solutions and several cases are examined in detail.  
In section 5 the stability of the solutions is considered. 
The mass  of the solutions 
and the boundary stress tensor are computed in Section 6.
We conclude with Section 7 where the results are compiled.

\section{GENERAL FRAMEWORK AND EQUATIONS OF MOTION}
\subsection{Basic ansatz}
Evolution of a minimally coupled complex scalar field $\Phi$ in $D$ 
dimensions is described by the action
(throughout this paper we will use units in which $c=\hbar=1$)
\begin{eqnarray}
\label{action}
S_D=-\int_{\mathcal{M}} d^D x    \left(\frac{\sqrt{-g_{\scriptscriptstyle D}}}{16\pi G_D}
(R-2\Lambda) +L_{\Phi}\right)
-\frac{1}{8\pi G_D}\int_{\partial\mathcal{M}} d^{D-1}x\sqrt{-h}K,
\end{eqnarray}
where the second term is the Hawking-Gibbons surface term \cite{Gibbons:1976ue}, $K$ is the trace 
of the extrinsic curvature for the boundary $\partial\mathcal{M}$ and $h$ is the induced 
metric of the boundary. Of course, this term does not affect the equations of motion 
but is important when discussing quantum properties of gravitating solutions.

In (\ref{action}) the lagrangian density of a complex self-gravitating scalar field reads
\begin{eqnarray}
\label{boson}
L_{\Phi}=-\sqrt{-g_{\scriptscriptstyle D}}\left( g^{i j}\Phi,_{i}^{\ast} \Phi_{j},
+ V(\Phi)\right),
\end{eqnarray}
where the asterisc denotes complex conjugate.
In this paper we consider only the  case $V(\Phi)=\mu^2 \Phi^{\ast}\Phi$,
where $\mu$ is the scalar field mass, without including 
a scalar self-interaction term. 
As found in \cite{Colpi:1986ye} for four dimensional asymptotically flat solutions, although the inclusion of 
a $\lambda |\Phi|^4$ term
drastically changes the value of the maximum mass and the corresponding 
critical central density of the boson star solutions, 
the qualitative features are essentially similar to the non-self interaction case.

The Lagrangian density (\ref{boson}) is invariant under a global phase rotation 
$\Phi \to \Phi e^{-i\alpha}$; 
that implies the existence of a conserved current 
\begin{eqnarray}
\label{J}
J^{k}=i g^{kl}
\left (\Phi^{\ast}_{,l}\Phi - \Phi_{,l}\Phi^{\ast} \right),
\end{eqnarray}
and an associated conserved charge, namely, the number of scalar particles
\begin{eqnarray}
N=\int d^{D-1} x \sqrt{-g_D} J^t.
\end{eqnarray}
The field equations are obtained by varying the action (\ref{action})  with respect 
to field variables $g_{ij}$ and $\Phi$
\begin{eqnarray}
\label{Einstein-eqs}
R_{ij}-\frac{1}{2}g_{ij}R+\Lambda g_{ij}&=&8\pi G_D T_{ij},
\\
\label{KG-eqs}
\nabla^2\Phi-\mu^2\Phi&=&0,
\end{eqnarray}
where the energy momentum tensor is defined by
\begin{eqnarray}
\label{Tij}
T_{ij}=\Phi^{\ast}_{,i}\Phi_{,j}+\Phi^{\ast}_{,j}\Phi_{,i}
-g_{ij}(g^{km}\Phi^{\ast}_{,k}\Phi_{,m}+\mu^2 |\Phi|^2).
\end{eqnarray}
As usual with the scalar field, the Klein-Gordon equation is redundant.

In this paper we assume a stationary ansatz $\Phi=\phi(r) e^{-i \omega t}$
describing a spherically symmetric bound state of the scalar field
with frequency $\omega$.
Hereafter, a prime denotes the derivative with respect to $r$ and an overdot $\partial/
\partial t$.
As we assume spherical symmetry it is convenient to use the metric form
\begin{eqnarray}
\label{metric}
ds^{2}=\frac{dr^2}{F(r)}+r^{2}d \Omega_{D-2}^2-F(r)e^{-2\delta(r)}dt^2,
\end{eqnarray}
where $d \Omega_{D-2}^2=\omega_{ab}dx^a dx^b$ is the line element on a unit $(D-2)$-dimensional sphere
and
\begin{eqnarray}
\label{F}
F(r)=1-\frac{2m(r)}{r^{D-3}}-\frac{2\Lambda r^2}{(D-2)(D-1)}.
\end{eqnarray}
The function $m(r)$ is related to the local mass-energy density up to some $D-$dependent factor.
Substituting in (\ref{Tij}) we find for the nonvanishing components of the energy-momentum tensor, 
\begin{eqnarray}
\nonumber
T_t^t&=&-\rho=-F \phi^{\prime 2}-\frac{\omega^2 e^{2\delta}}{F}\phi^2-\mu^2 \phi^2,
\\
\nonumber
T_r^r&=&P_r=F \phi^{\prime 2}+\frac{\omega^2 e^{2\delta}}{F}\phi^2-\mu^2 \phi^2,
\end{eqnarray}
for the energy density and the radial pressure while for the $(D-2)$ angular components we find
\begin{eqnarray}
\nonumber
T_{\theta}^{\theta}=P_t=-F \phi^{\prime 2}+\frac{\omega^2 e^{2\delta}}{F}\phi^2-\mu^2 \phi^2.
\end{eqnarray}
We remark that the radial component of the pressure $P_r$ 
and tangential components $P_t$ are different,
precluding a macroscopic description of these configurations in terms
of an effective perfect fluid energy-momentum tensor.
\subsection{Reduced action and virial relations}
If one is only interested in spherically symmetric solutions it is much simple 
to work with a reduced action where this symmetry of the spacetime is factored out.
Expressing  the curvature  scalar $R$ in terms of the metric  function  $m(r)$ and
$\delta(r)$, we obtain the following  expression for the effective action of our static
spherically symmetric system
\begin{eqnarray}
\label{Leff}
S(m,\delta,\phi)=\int_{0}^{\infty} \left(\frac{(D-2)}{8 \pi G_D}m'e^{-\delta}
-r^{D-2}e^{-\delta}\Big(F\phi'^2
-\frac{e^{2\delta}\omega^2 \phi^2}{F}+V(\phi) \Big) \right) dr.
\end{eqnarray}
This form of the reduced action allow to derive an useful virial relation 
by using a scaling technique.
In this way it is possible to better understand the reason for the existence of 
nontrivial solutions and to provide nonexistence theorems.

We will use an approach proposed by Heusler in \cite{Heusler:1996ft}. 
Let us assume the existence of a solution $ m(r), \delta(r), \phi(r)$
 with suitable boundary conditions at the origin and at infinity. 
 Then each member of the 1-parameter family
\begin{equation}
m_\lambda(r) \equiv m(\lambda r), \ 
\delta_{\lambda} (r) \equiv \delta (\lambda r), \ 
\phi_{\lambda} (r) \equiv \phi (\lambda r)
\end{equation}  
assumes the same boundary values at $r=0$ and $r=\infty$, and the action 
$S_{\lambda} \equiv S[m_{\lambda}, \delta_{\lambda},\phi_{\lambda}]$
must have a critical point at $\lambda=1$, $[dS/d\lambda]_{\lambda=1}=0$.
Thus  we obtain the virial relation valid for a $D-$dimensional spacetime
\begin{eqnarray}
\label{v1}
\int_0^{\infty}dr_{}~r^{D-2} e^{-\delta} \Bigg(
(D-3)\left(1+\frac{4m}{r^{D-3}}-\frac{2\Lambda r^2}{(D-2)(D-3)}\right)\phi'^2
+(D-1)V(\phi)
\nonumber
\\
-\left(D-1-\frac{4(D-2)m}{r^{D-3}}-\frac{2(D-3)\Lambda r^2}{(D-1)(D-2)}\right)
\frac{\omega^2 e^{2\delta}\phi^2}{F^2}
\Bigg)=0.
\end{eqnarray}  
By using this relation we can also discuss whether the cosmological constant 
can support a real scalar field.
In asymptotically flat spacetimes, a well-known result implies the absence of 
scalar solitons. 
It can be seen from (\ref{v1}) that in any dimension $D \geq 3$ 
there are no static, spherically symmetric scalar solutions 
with negative cosmological constant and a positive potential, independently of whether 
or not gravity is taken into account.

Also, there are different proofs of the no-hair theorem for spherically
symmetric scalar fields (see \cite{Heusler:1996ft} for a set of references).  
The above scaling argument does not exclude black hole
solutions with complex scalar hair 
(to this end, one has to replace the lower boundary 
in the action integral (\ref{Leff}) by the horizon distance $r_h$).
However, by using another sort of argument, 
we will prove in Section 3 the absence of such solutions, 
which clearly indicates the limits of scaling techniques.

For $V(\phi)=\mu^2 \phi^2$, the  relation (\ref{v1}) can be simplified
by using the Klein-Gordon equation written in the form
\begin{eqnarray}
\frac{1}{2}(r^{D-2} e^{-\delta}F (\phi^2)')'
=r^{D-2}\left(F\phi^{\prime 2}+\mu^2 \phi^2 -
\phi^2 \frac{\omega^2  e^{2 \delta}}{F} \right).
\end{eqnarray} 
Integrating the above relation and making use of the boundary conditions
from Section (2.3), we find
\begin{eqnarray}
\int_{0}^{\infty} dr~ r^{D-2} e^{-\delta} \left( F \phi^{\prime 2}+\mu^2 \phi^2-
\phi^2 \frac{\omega^2  e^{2 \delta}}{F} \right)=0,
\end{eqnarray} 
which, together with (\ref{v1}) gives the simple relation
\begin{eqnarray}
\label{v2}
\int_{0}^{\infty} dr~r^{D-2} e^{-\delta} \left(
(1-\frac{m(3D-7)}{r^{D-3}})\phi^{\prime 2}+\frac{\omega^2 e^{2 \delta}}{F^2}
(\frac{2\Lambda r^2}{(D-1)(D-2)}-\frac{m(D-3)}{r^{D-3}})\phi^2 \right)=0.
\end{eqnarray} 
In the above relation, the factor in front of $\phi^2$ has a negative sign.
Thus, the existence of boson star solutions is supported by the factor 
$(1-m(3D-7)/r^{D-3})$ in front of $\phi^{\prime 2}$, which has no definite sign.

The relation (\ref{v2}) particularized for $D=3$ 
\begin{eqnarray}
\label{v3}
\int_{0}^{\infty} dr~r  e^{-\delta} \left(
(1- 2m) \phi^{\prime 2}+\frac{\omega^2 e^{2 \delta}\Lambda r^2 }{F^2}
\phi^2 \right)=0.
\end{eqnarray}
can be used to exclude the existence
of three-dimensional asymptotically flat boson stars.

\subsection{Field equations and boundary conditions}
To perform numerical computations and order-of-magnitude estimations,
it is useful to have a new set of dimensionless variables. 
Thus, we perform the rescalings $r \to r/\mu $,
$\phi \to \sqrt{\frac{(D-2)}{16\pi G_D}} \phi,~m \to m/\mu^{D-3} $ and $\Lambda \to \Lambda/\mu^2$, 
while the factor $\omega/\mu$ is absorbed into the definition
of the metric function $\delta$.

Then we find the field equations
\\
\begin{eqnarray}
\label{eqm}
m'&=&\frac{r^{D-2}}{2}\left(F\phi'^2+\phi^2+\frac{e^{2\delta}\phi^2}{F}\right),
\\
\label{eqdelta}
(e^{-\delta})'&=&r\left(e^{-\delta}\phi'^2+ e^{\delta}\frac{\phi^2}{F^2} \right),
\\
\label{eqfi}
(r^{D-2}e^{-\delta}F\phi')'&=&r^{D-2}e^{-\delta}\phi
\left( 1-\frac{e^{2\delta}}{F} \right).
\end{eqnarray}
Taking into account the explicit form for the metric and the scalar field 
we find the particle number
\begin{eqnarray}
\label{number}
N=\frac{4\pi^{(D-1)/2}}{\Gamma(\frac{D-1}{2})}
\int_{0}^{\infty}dr~ \phi^2 r^{D-2}\frac{e^{\delta}}{F}.
\end{eqnarray}
Similar to the $\Lambda=0$ case, we can define a star radius
\begin{eqnarray}
\label{R}
R=\frac{1}{N} \int~d^{D-1} x~r J^t \sqrt{-g_D}
=\frac{4\pi^{(D-1)/2}}{\Gamma(\frac{D-1}{2})N}
\int_{0}^{\infty}dr~ \phi^2 r^{D-1}\frac{e^{\delta}}{F}.
\end{eqnarray}
Following the standard analysis, we can predict the 
boundary conditions and some general features of the finite-energy solutions.
The boundary conditions for the system are the following.
The non-singularity at the center of the star requires that $\phi(0)=\varphi_0,~\phi'(0)=0,~m(0)=0$.
We find also that the boundary conditions at infinity 
obtained for an asymptotically flat spacetime remain valid, with $\phi=0$ as the  only
acceptable value, and also that
\begin{eqnarray}
\lim_{r \to \infty} \phi'^2 \sim O(1/r^{D+2+\epsilon}),
\end{eqnarray}
where $\epsilon$ is a small positive quantity.

The formal power series describing the above boundary conditions at $r=0$ is
\begin{eqnarray}
\label{expansion}
\phi(r)&=&\varphi_0+\frac{\varphi_0(1-e^{2\delta_0})}{2(D-1)} r^2+O(r^3),
\\
\label{expansion-m}
m(r)&= & \frac{\varphi_0^2(1+e^{2\delta_0})}{2(D-1)} r^{D-1}+O(r^D),
\\
\delta(r)&=&\delta_0-\frac{1}{2}e^{2\delta_0}\varphi_0^2 r^2+O(r^3).
\end{eqnarray}
The dimensionless value of the scalar field at the origin $\varphi_0$ 
is also used as a parametrization of the central density of the configuration.
For $\Lambda=0$, boson star solutions are characterized 
by an exponential decay of the scalar field,
for which a mass term in the potential is responsible. 
The behaviour for $\Lambda<0$ is different:
the analysis of the field equations as $r\to\infty$ gives
\begin{eqnarray}
\label{asymptotics}
\nonumber
\phi(r)&\sim &\hat{\phi}_0r^c+\dots,
\\
m(r)&\sim & M+\frac{(k^2 c^2 +1)\hat{\phi}_0^2}{2(2c+D-1)}r^{2c+D-1}+\dots,
\\
\nonumber
e^{-\delta(r)}&\sim & 1+\frac{c\hat{\phi}_0^2}{2}r^{2c}+\dots,
\end{eqnarray}
where $c=-\frac{1}{2}\left( D-1+\sqrt{(D-1)^2+4/k^2}\right),~k^2
=-2\Lambda/(D-2)(D-1)$ and $\hat{\phi}_0$ is a constant.
Thus, the cosmological constant implies a complicated power decay at infinity,
rather than an exponential one as found in
an asymptotically flat space.

We will prove in Section (6) that the constant $M$ in the above relations

is the $ADM$ mass of a boson star up to a $D$-dependent factor.
However, in the discussion of numerical solutions we will refer to $M$ to as the mass of the solutions.

\section{ A NO-HAIR THEOREM}
In general, when a theory allows regular configurations,
it also allows black hole solutions.
However, this is not the case for the action discussed in this paper.
In four dimensions with $\Lambda=0$, Pe\~na and Sudarsky  have proven that there 
are no black hole analogues of the regular
boson star configurations \cite{Pena:1997cy}.
Therefore, the collapse of a boson star results in a trivial Schwarzschild
black hole.
A crucial point in their proof was the asumption of asymptotic flatness.
However, in the last years it has been realized that the introduction of a
(negative) cosmological constant may allow for configurations forbidden   
in the asymptotically flat case
(see, for example, the existence of dyonic regular and black hole solutions
in EYM theory with $\Lambda<0$).
The dimensionality of spacetime may also affect the existence and the properties of solutions
in a given theory.
A nontrivial example is the gravitating SU(2) theory, which presents spherically symmetric
particle-like solutions
in $D=4$ \cite{Bartnik:1988am} without counterparts
in $D=3$ \cite{Deser:1984fk} and $D=5$ \cite{Volkov:2002tb} dimensions.

In this section we prove the following straightforward generalization of the
Pe\~na-Sudarsky theorem for $D\geq 3$ 
and $\Lambda \leq 0$.

\emph{Theorem:}\\
A $D-$dimensional, static, spherically symmetric, AAdS black hole spacetime, 
with regular event horizon, satisfying Einstein's equations with
the matter fields fulfilling the weak energy condition (WEC) and its
energy-momentum satisfying the condition
\begin{eqnarray}
\label{condition}
T_{\theta}^{\theta} \leq T_r^r,
\end{eqnarray}
is necesarily trivial ($i.e.$ $T_{i}^{j}=0 $
and the only black hole is Schwarzschild-AdS spacetime).

\emph{Proof.:}~~~
Since our arguments are very similar to \cite{Pena:1997cy}, 
we present only the main steps. 
From the ($rr$) and ($tt$) Einstein equations we can derive the following relations
\begin{eqnarray}
F'&=&\frac{16 \pi r G_D}{D-2}T_t^t-\frac{2r\Lambda}{D-2}
+\frac{D-3}{r}(1-F),
\\
\delta'&=&\frac{8\pi r G_D}{(D-2)F}(T_t^t-T_r^r)
\end{eqnarray}
By using these relations and the conservation of 
the energy-momentum tensor $T^k_{i;k}=0$ we find
\begin{eqnarray}
\label{th1}
e^{\delta}(e^{-\delta} T^r_r)'=\frac{1}{2Fr}\left[
(D-3)(1-F-\frac{2\Lambda r^2}{(D-2)(D-3)})(T_t^t-T^r_r)+
2F(D-2)(T_{\theta}^{\theta}-T_r^r)
\right].
\end{eqnarray}
The WEC states that the energy density $\rho\equiv-T_t^t $ is semipositive definite
and $T_r^r,T_{\theta}^{\theta} \geq T_t^t$.
Thus, this condition and the assumption (\ref{condition}) implies that 
$e^{-\delta} T^r_r$ is a nonincreasing function of $r$ and
$e^{-\delta} T^r_r (r) \leq e^{-\delta} T^r_r (r_h)$.
Here $r_h$ is the event horizon radius, corresponding to a zero of the function $F$.
However, by expressing (\ref{th1}) in terms of the proper radial distance
$dx=\frac{dr}{\sqrt{F}}$ we find
\begin{eqnarray}
\label{th2}
e^{\delta}\frac{d}{dx}(e^{-\delta} T^r_r)&=&\frac{1}{2\sqrt{F} r}\left(
(D-3)(1-\frac{2\Lambda r^2}{(D-2)(D-3)})(T_t^t-T^r_r)
\right)
\\
&+&
\frac{\sqrt{F}}{2r}
\Big (
2(D-2)(T_{\theta}^{\theta}-T_r^r) -(D-3)(T_t^t-T^r_r)
 \Big).
\end{eqnarray}
Since the left hand side of this equation is finite as $r \to r_h$,
we find $T^r_r(r_h)=T_t^t(r_h)=-\rho(r_h)<0$.
However $e^{-\delta} T^r_r(\infty)=0$, so we conclude $T_r^r(r)=0$.
The WEC plus condition (\ref{condition}) implies that $T_{\theta}^{\theta}=T_t^t=0$ also.
The only black hole solution of Einstein field equations is thus 
the Schwarzschild-AdS spacetime.

It can easily be seen that a complex scalar field with the lagrangian density
$(\ref{boson})$ and $\Phi=\phi(r) e^{-i \omega t}$
satisfies the conditions of the theorem above as long as $V(\Phi)>0$.

By using similar arguments we can also 
exclude the nonexistence of topological black holes.
These are black holes with $\Lambda<0$ for which the topology of the horizon 
is an arbitrary genus Riemann surface, the
$(D-2)$-sphere being replaced by a $(D-2)$-dimensional space
of negative or vanishing curvature (see \cite{Lemos:2000un} for a review).

Applying the same arguments in the case of a positive cosmological constant  
leaves open the question of the existence.
We find only
that the scalar field and the energy momentum tensor should vanish 
on the cosmological event horizon \cite{Cai:1997ij} while $T_r^r$ 
is a strictly increasing quantity.

\section{NUMERICAL SOLUTION}

While an analytical solutions to the coupled nonlinear equations  
(\ref{eqm})-(\ref{eqfi}) 
appears to be intractable for every dimensions,
the resulting system has to be solved 
numerically.
In this way we find that the boson stars with negative cosmological constant 
may exist in any dimension $D \geq 3$.

A complete analysis of the complex correlation between the two parameters
of the theory ($\varphi_0, \Lambda$) for a given spacetime dimension 
is beyond the purposes of this paper.
To compare numerically the results with those known in the asymptotically flat case we 
focused on solutions with $D=3,~4,~5$ and have varied the parameter $\varphi_0$ for a 
set of $\Lambda$.

For each choice of $\varphi_0$, the eqs. (\ref{eqm})-(\ref{eqfi})
with the initial conditions (\ref{expansion}) 
have a solution satisfying the boundary conditions  (\ref{asymptotics})
only when $e^{\delta}(r=0)$ takes on certain values.
Different values of $\delta_0$ corresponds to different number of nodes of the scalar
field. 
We follow the usual approach and, by using a standard ordinary  differential  
equation solver, we
evaluate  the  initial conditions at $r=10^{-3}$ for global tolerance    
$10^{-12}$, adjusting for fixed shooting parameter and integrating  towards  $r\to\infty$.
We solve the equations and find 
the functions $m,~\delta$ and $\phi$ and also the values of $M,~N,~R$.
The energy of these solutions is always concentrated in a small region.

Only the zero-node solutions will be given here.
However, we found that similar to the $D=4,~\Lambda=0 $ case \cite{Mielke:1980sa}, 
$\phi$ can have nodes, giving rise to excited states.

 
\subsection{Boson stars in four dimensions} 
We start by discussing the better known case $D=4$.
The behaviour of $M,~N$ as a function of $\phi(0)$ is well known for the
$\Lambda=0$ case; the mass and particle number rise with the increasing 
of $\phi(0)$ to a maximum of $M_{max}=0.633~M_{Pl}^2/\mu$ and 
$N_{max}=0.653~M_{Pl}^2/\mu^2$. After this $M$ and $N$ drop, 
oscillate a bit and finally approach constant values independent of $\phi(0)$.

From $Figures$ 1b,~4b it is clear that the introduction of $\Lambda<0$ does not change 
this behaviour qualitatively.
The location of maximum shifts to higher values of $\phi(0)$
while the maximum mass and particle number decreases with the value of 
the cosmological constant.
However, the general properties of the solutions are the same as for the $\Lambda=0$ case.
The results for $\Lambda=0$ agrees with previous results in the literature.

As a general feature,
 we have noticed a decresing of the maximal 
allowed value of the parameters $N, M$ and a smaller ADM mass for the same central density as
compared with $\Lambda=0$ case. 
For example,  when $\Lambda=-0.05$ we have found that
$M_{max}=0.480$ and $N_{max}=0.435$, while for $\Lambda=-1$, 
these values are $M_{max}=0.239$, $N_{max}=0.121$.
Different values for the shooting parameter $\delta_0$ are found.

From Figure 2b it is evident that the characteristic boson-star 
masses decrease with increasing $|\Lambda|$.
We can qualitatively understand this fact in the following way.
Adding a negative cosmological constant corresponds to adding an attractive force. 
The Heisenberg uncertainty principles requires that a quantum state confined 
into a region of characteristic radius $R$
has a typical boson momentum $p \sim 1/R$.
Since for a marginally relativistic boson star $p \sim \mu$ we find $R \sim 1/\mu$.
The order of magnitude of the critical mass for the formation of a black hole 
can be estimated by comparing $R$ with the Schwarzschild radius $r_s$.
Since the relation between  Schwarzschild black hole mass and event horizon radius is
$2M=r_h(1-\Lambda/3r_h^2)^{-1} \sim r_h (1+\Lambda/3r_h^2)$, we find a critical mass
 $M_c \sim  (1+\Lambda/3\mu^2)M_{Pl}^2/\mu$, always smaller than the asymptotically flat value.
 
 Note that, again, for a finite value of the central density, the binding energy
 of the configurations $E_b \equiv M-N$ becomes positive, signaling the existence of configurations
 with excess energy.
 This property express a global instability against dispersion of particles to infinity.
 The excess energy is translated into kinetic energy of the free particles at infinity.
 
 
\subsection{Boson stars in five dimensions} 
We integrated the equations of motion for a number of dimensions greater than four. 
We find that the general picture we presented for $D=4$ remains always   
valid for higher dimensional spherically symmetric bosons stars.

The parameter $M$ rises quickly with the initial value of the scalar 
field $\phi(0)$ at a maximum,
afterwards drops, and has several local minima and maxima until it approaches an asymptotic
value for large $\phi(0)$. A similar behaviour is found for the particle number.
The location of the first peak again shifts 
to higher values of $\phi(0)$ for a negative $\Lambda$.
Curiously, the five-dimensional asymptotically flat solutions present this peak 
for very small values of $\varphi_0$.

The results for three diferent values of $\phi(0)$ and a varying
 $\Lambda$ are presented in $Figures$ 2c, 3c.
 As expected, the values of $(M,~N)$ for a given $\varphi_0$ depends on the value of $\Lambda$.
For a nonzero $\Lambda$ it is necesary to establish new (and smaller) limiting values for 
the maximal mass and maximal particle number. 
For example, an asymptotically flat  
five-dimensional solution has $M_{max}=3.803$ and $N_{max}=3.792$; 
when $\Lambda=-0.01$ we have found that
$M_{max}=1.764$ and $N_{max}=1.667$, while for $\Lambda=-0.1$, 
these values are $M_{max}=0.893$, $N_{max}=0.719$.
We can heuristically understand this fact by using the arguments presented for $D=4$.
Typical results of the numerical  integration are  presented in $Figures$ 1c,4c.

Note that again for a finite value of the central density 
the binding energy of the configurations $E_b\equiv M-N\mu$ becomes 
positive, signaling the existence of configurations
with excess energy. 
 
\subsection{Boson stars in three dimensions} 
The (2+1) dimensional boson stars are a rather special case.
Three dimensional gravity provided us 
with many important clues about higher dimensional physics.
It help that this theory with a negative cosmological constant $\Lambda$   
has non-trivial solutions, such as the BTZ
black-hole spacetime \cite{Banados:wn}, which provide important testing ground
for quantum gravity and AdS/CFT correspondence.
Many other types of $3D$ regular and black hole solutions with a negative cosmological constant 
have also been found by coupling matter fields to gravity
in different way.
(2+1)-dimensional stars with negative cosmological constant have been studied in
\cite{CZ,Lubo:1998ue}.
We notice also the existence of interesting BTZ-like solutions in a class of 3D gravity models
in which the cosmological constant is induced \cite{Mielke:1991nn}.

Exact boson star solutions with $\Lambda<0$ (in the limit of large self interaction) 
are also known to exist \cite{Sakamoto:1998hq}.

The line element  (\ref{metric}) reads in this case 
\begin{eqnarray}
\label{3D}
ds^{2}=\frac{dr^2}{1-2m(r)-\Lambda r^2}+r^{2}d \theta ^2-(1-2m(r)-\Lambda r^2)e^{-2\delta}dt^2, 
\end{eqnarray}
where the angular variable $\theta$ is assumed to vary between $0$ and $2 \pi$.
At the first sight, it may be that the boundary condition (\ref{expansion-m})
is too strong and we have the freedom to add a constant to the function $m(r)$ \cite{Sakamoto:1998hq}.
However, if we assume that this metric posseses a symmetry axis 
(located at $r=0$), and has no conical singularities,
we have to impose the condition $\lim_{r \to 0} m(r)=0$.
Asymptotically, the line element (\ref{3D}) approaches the BTZ metric
\begin{eqnarray}
\label{BTZ}
ds^{2}=\frac{dr^2}{-\tilde{M}-\Lambda r^2}+r^{2}d \theta ^2-(-\tilde{M}-\Lambda r^2)e^{-2\delta}dt^2, 
\end{eqnarray}
with $\tilde{M}=-1+2m(\infty)$.
We will find in Section 6 that the interpretation of the function $m(r)$ and of its limit 
as $r \to \infty$ is more subtle .

 We have integrated the field equations with the initial conditions (\ref{expansion}) 
for a number of $\Lambda<0$ and have found a different picture 
as compared to the case $D>3$.
The solution with $\varphi_0=0$ corresponds to the global $AdS_3$. 
For a given $\Lambda$, we find nontrivial solutions up to a maximal value of 
$\varphi_0$, where
the numerical iteration diverges.
A divergent result is obtained also in the limit $\Lambda \to 0$.
The results of the numerical integration both for $M$ and $N$ and three values of $\Lambda$ 
are presented
in $Figures$ 1a,~4a.
In $Figures$ 2a, 3a we present the variation of $M$ and $N$ with $\Lambda$ for 
fixed values of the central density. 

We noticed that, for every $\Lambda,\phi_0$, the asymptotic value of the   
function $m(r)$ is smaller that 1/2.
However, the maximal allowed value of the particle number can be higher than 1/2 
(for small enough values of $|\Lambda|$).

Further details on these  solutions as well as rotating 3D configurations 
will be presented elsewhere.

\section{STABILITY OF SOLUTIONS VIA LINEAR PERTURBATION THEORY}
An important physical question when discussing selfgravitating 
configurations is whether these solutions are stable.
For a boson star, the discussion of gravitational equilibrium is more 
complicated since we cannot apply the well-known
stability theorems for fluid stars.
When the spacetime is not asymptotically flat
the stability analysis can also be a quite involved and subtle problem.

A number of authors discussed the dynamical stability of four dimensional $\Lambda=0$ boson 
stars by means of linear perturbation theory \cite{Gleiser:1988rq}-\cite{Lee:av}.
This lead to an eingenvalue problem, which is of Sturm-Liouville type and 
which determines the normal modes of the radial oscillations and their eingenvalues 
$\chi^2$.
The sign of the lowest eingenvalue $\chi_0^2$ is crucial; if $\chi_0^2>0$ then the star is stable,
whereas it is unstable.
In particular Gleiser and Watkins used a powerful method to discuse the stability of 
asymptotically flat boson stars, by studying the frequency spectrum
of a class of radial perturbations \cite{Gleiser:1989ih}. 
They have shown
that the transition from stability to instability occurs always at 
the critical points of mass (or charge) against the value of the scalar field at the origin.
A more powerful method based on catastrophe theory was also proposed \cite{Kusmartsev:cr}, 
with similar conclusions.

Since the general picture we find for $D \geq 4$ is similar to the well known asymptotically flat solutions,
we expect to find similar results for the stability question.
Therefore we conjecture that $M_{max}$ represents the boundary between 
stable and unstable gravitational equilibrium.

The method proposed by Gleiser and Watkins  \cite{Gleiser:1989ih} 
will be used here to discuss 
the  stability of $D$ dimensional solutions with $\Lambda \leq 0$. 
Since our approach is practically similar to that presented in \cite{Gleiser:1989ih}, 
we will present here the main steps only, emphasizing the points where 
the different asymptotic structure of spacetime is  essential.
We find that, for any dimension, the perturbation equations reduce to a set of two coupled differential equations.

We consider the situation where the equilibrium configuration is perturbed in a way
such that the spherical symmetry is still preserved.
These perturbations will give rise to motions in the radial direction.

To conform with previous works in $D=4$, we use in this section the following parametrization
for the spherically symmetric line element
\begin{eqnarray}
\label{mp}
ds^2=e^{\lambda}dr^2+r^2 d\Omega^2_{D-2}-e^{\nu}dt^2,
\end{eqnarray}
where $\nu$ and $\lambda$ are functions of $r$ and $t$ only.

The equations governing the small perturbations are obtained by expanding all functions to first order
and then by liniarizing the equations (\ref{Einstein-eqs})-(\ref{KG-eqs}).
We  write therefore
\begin{equation}
\begin{array}{l}
\lambda(r,t)=\lambda_0(r)+\delta \lambda(r,t),
\\
\nu(r,t)=\nu_0(r)+\delta \nu(r,t),
\end{array}
\end{equation}
where $\delta \lambda(r,t),~\delta \nu(r,t)$ are related to the perturbations of  
$m,~\delta$.
The scalar field is written as
\begin{eqnarray}
\phi(r,t)=\big ( \phi_0(r) +\delta \phi(r,t) \big)e^{-i \omega t}.
\end{eqnarray}
Similar to the $\Lambda=0$ four dimensional case, it is possible to reduce the full system 
of equations  
to only two coupled equations in two unknowns.
If we choose to obtain these equations for $\delta \phi$, 
it is convenient to introduce the notations 
$\delta \phi(r,t)=f_1(r,t)+i \phi_0(r) \dot{g}(r,t)$ (in some cases,
the analytically equivalent set of equations for $\delta \lambda,~f_1$ may be more suited
in the numerical computation).
For the equilibrium configurations the metric functions
are time independent, 
$e^{\lambda_0}=1/F,~e^{\nu_0}=e^{-2\delta}F$ and the scalar field
is $\phi(r,t)=\phi_0(r)e^{-i \omega t}$.

For this scalar field ansatz, the linearized $(r,t)$ Einstein equation can be trivially 
integrated in time giving
\begin{eqnarray}
\label{p0}
\delta \lambda =\frac{2r}{D-2} 16 \pi G_D (\phi_0'f_1-\omega\phi_0^2 g')
.
\end{eqnarray}
The $(r,r)$ and $(t,t)$ Einstein equations give two more equations for the metric perturbations
\begin{eqnarray}
\label{p1}
&&-\frac{(D-2)}{2r^2} \left[(D-3) e^{-\lambda_0} \delta \lambda +re^{-\lambda_0} \delta \lambda'
-re^{-\lambda_0} \lambda_{0}'  \delta \lambda\right]
=8 \pi G_D \Big[ e^{-\nu_0} \delta \nu \omega^2 \phi_0^2 
\\
\nonumber
&&+e^{-\lambda_0} \delta \lambda \phi^2_{0,r}
-2e^{-\nu_0} \omega \phi_0 (\omega f_1-\phi_0\ddot{g})
-2e^{-\lambda_0}\phi'_{0}f'_{1}-2\mu^2 \phi_0 f_1 \Big],
\\
\label{p2}
&&\delta \nu'=(D-3)\frac{\delta \lambda}{r}+\nu_0'\delta \lambda +\frac{2r}{D-2} 
8\pi G_D e^{\lambda_0}
\Big[-e^{-\lambda_0} \phi_0^{\prime 2}\delta \lambda+2 e^{-\lambda_0} \phi_0' f_1'
-2\mu^2 \phi_0 f_1,
\\
\nonumber
&&-e^{-\nu_0} \omega^2 \phi_0^2 \delta \nu
+2 e^{-\nu_0} \omega \phi_0 (\omega f_1 -\phi_0 \ddot{g})\Big],
\end{eqnarray}
while the linearized Klein-Gordon equation for $f_1$ and $g$ are
\begin{eqnarray}
\label{p3}
&&f_{1}''+f'_{1}\left(\frac{D-2}{r}+\frac{1}{2}(\nu_0'-\lambda_0')\right)
-e^{\lambda_0-\nu_0}\ddot{f}_1
+f_1e^{\lambda_0}(\omega^2 e^{-\nu_0}-\mu^2)-2\omega e^{\lambda_0 -\nu_0} \phi_0 \ddot{g}
\\
\nonumber
&&+\frac{1}{2}(\delta \nu '-\delta \lambda')
+e^{\lambda_0-\nu_0}\omega^2 \phi_0 (\delta_\lambda-\delta \nu)-
\mu^2 e^{\lambda_0} \phi_0 \delta \lambda=0,
\\
\label{p4}
&&g''+g'\Big[\frac{D-2}{r}+\frac{1}{2}(\nu_0'-\lambda_0')+2\frac{\phi_0'}{\phi_0}\Big]
-\ddot{g} e^{\lambda_0 -\nu_0}+2\omega e^{\lambda_0 -\nu_0} \frac{f_1}{\phi_0}
+\frac{1}{2}\omega e^{\lambda_0 -\nu_0}(\delta_\lambda-\delta \nu)=0.
\end{eqnarray}
The procedure now is simple. Using eq. (\ref{p0}), it is possible to eliminate
$\delta \lambda$ from (\ref{p3}). 
An expression for $\delta \nu$ is obtained for eq. (\ref{p4}).
We can also differentiate eq. (\ref{p4}) with respect to $r$ and eliminate
$\delta \nu'$ by using eq. (\ref{p2}).
In this way we find
\begin{eqnarray}
\label{p1-f}
f_1''+f_1'\left( \frac{D-2}{r}+\frac{1}{2}(\nu_0'-\lambda_0')\right)
-\ddot{f_1} e^{\lambda_0-\nu_0}+C_1(r)f_1
-2\omega \phi_0 g''-C_3(r)g'
\\
\nonumber
-\omega g' \left(2\phi_0'+\phi_0 
(\frac{2D-4)}{r}+ \nu_0'-\lambda_0' )\right)=0,
\\
\label{p2-f}
g'''+g''\big[ \frac{D-2}{r}+\frac{3}{2}(\nu_0'-\lambda_0')+\frac{\phi'_0}{\phi_0}\big]
-e^{\lambda_0-\nu_0}\ddot{g}'+2e^{\lambda_0-\nu_0}\frac{\omega}{\phi_0}f_1'
-C_2(r) g'-\frac{e^{\lambda_0-\nu_0}}{\phi_0^2}C_3(r) f_1=0,
\end{eqnarray}
where
\begin{eqnarray}
\label{Ci}
\nonumber
 C_1&=&-e^{\lambda_0}(\mu^2+3\omega^2 e^{-\nu_0})
+\frac{16 \pi G_{D}}{D-2} r \phi_0^{\prime 2} 
(\frac{2D-6}{r}+\nu_0'-\lambda_0')
-\frac{64 \pi G_{D}}{D-2} r e^{\lambda_0} \phi_0' \phi_0 \mu^2,
\\
C_2&=&(D-2) \frac{(\lambda_0'-\nu_0')}{r}+\frac{D-2}{r^2}-
\frac{1}{2}(\nu_0^{''}-\lambda_0^{''})+
2(\frac{\phi_0'}{\phi_0})^2
-2\frac{\phi_0^{''}}{\phi_0}
\\
\nonumber
&{}&+\frac{1}{2}(\lambda_0'-\nu_0')(\nu_0'-\lambda_0'+4\frac{\phi_0'}{\phi_0})
-\frac{16 \pi G_{D}}{D-2} re^{\lambda_0-\nu_0}\omega^2 \phi_0^2 
(\nu_0'-\lambda_0'+\frac{2D-6}{r}),
\\
\nonumber
C_3&=&\omega\left(2\phi_0'+\frac{16 \pi G_{D}}{D-2}r \phi_0^2 \phi_0'
(\frac{2D-6}{r}+\nu_0'-\lambda_0')
-\frac{32 \pi G_{D}}{D-2}r\phi_0^3\mu^2 e^{\lambda_0} \right).
\end{eqnarray}
We consider only radial perturbations, which conserve the total
particle number $N$, as given by (\ref{number}), from which we can compute $\delta N$.
Remarkably, we find that always the integrant of $\delta N$ is a total derivative
with respect to $r$
\begin{eqnarray}
\delta N=\frac{8\pi^{(D-1)/2}}{\Gamma(\frac{D-1}{2})}\phi_0^2 g' r^{D-2}
e^{\frac{\nu_0-\lambda_0}{2}}\Big|_0^{\infty}.
\end{eqnarray}
The appropriate boundary conditions for $r \to \infty$ are $f_1 \to 0,
~g'r^{D+2c} \to 0$ while $f_1=$const., $r^2g' \to 0$ at the origin.

Now we assume that all perturbations are of the 
form $e^{i\chi t}$, where $\chi$ is the characteristic frequency to be determined.
The self-adjoint system of coupled equation (\ref{p1-f}) and (\ref{p2-f}) 
can be written in the compact form
\begin{eqnarray}
L_{ij}f_j=\chi^2 M_{ij}f_j, ~~~i,j=1,2
\end{eqnarray} 
 with $f_2=g'$ and
\begin{displaymath}
L_{ij}=\left( \begin{array}{cc}
-\frac{d}{dr} G_1\frac{d}{dr} +G_1C_1
&2\omega \frac{d}{dr} G_1\phi_0 +G_1C_3\\
-2 \omega G_1 \phi_0 \frac{d}{dr}+G_1 C_3 & -\frac{d}{dr} G_2\frac{d}{dr} +G_2C_2
\end{array}
\right),~~~~
M_{ij}=e^{\lambda_0-\nu_0}\left( \begin{array}{cc}
G_1 &0\\0&G_2 
\end{array}
\right),
\end{displaymath}
where $G_1=r^{D-2}e^{(\nu_0-\lambda_0)/2},~G_2=r^{D-2}\phi_0^2e^{3(\nu_0-\lambda_0)/2}$,
which is the required "pulsation equation".
This sytem defines a characteristic value problem
for $\chi^2$ which must be real
(note the close analogy with the $D=4,\Lambda=0$ case).
With the above boundary conditions, it can be easily verified that both
$L_{ij}$ and $M_{ij}$ satisfy the condition
\begin{eqnarray}
\nonumber
\int_{0}^{\infty}dr \eta_i L_{ij} \xi_j=\int_{0}^{\infty}dr \xi_i L_{ij} \eta_j,
\end{eqnarray}
and similarly for $M_{ij}$. The equations  (\ref{p1-f}) and (\ref{p2-f}) 
can also be derived from a variational principle.
Eigenfunctions related to different eigenvalues 
satisfy also an orthogonality relation.
The eigenvalues $\chi^2$ form a discrete sequence 
$\chi_0^2 \leq  \chi_1^2\leq  \chi_2^2 \dots$.
Thus, if the fundamental mode is stable  ($\chi_0^2\geq 0$), then all radial modes are stable.
Also, if the boson star is radially unstable,
the fastest growing instability will be the fundamental one.
We are interested in finding the values of $\phi_0(0)$ 
for which the equilibrium configurations are stable.
As extensive discussed in \cite{Gleiser:1989ih}, 
we can find some insight in this problem by considering static perturbation.
If for a given equilibrium configuration a radial mode changes its stability property 
(therefore its corresponding eingenvalues $\chi_0^2$ passes through zero), then there will exist
infinitesimally nearby equilibrium configurations with the same total
mass and total particle number.
Hence, if $\chi^2$ goes through zero we have $dM/d\phi_0(0)=dN/d\phi_0(0)=0$.

Also, if we begin with an equilibrium configuration with $\phi_0$, the perturbed fields will
describe another equilibrium configuration with $\phi_0+\delta \phi_0$.
However, the perturbed configurations must have the same charge as the equibrium configurations.
Therefore, zero frequency perturbations will exist if and only if there exist 
two neighboring configurations with the same charge.
This occurs at the extremum of $N$, where $dN/d\phi_0(0)=0$. 
This is a very general argument and aplies for every dimension.

For all discussed cases except $D=3$, the behaviour of the mass and particle number, 
parametrized by $\varphi_0$ suggest that the solutions
with $\varphi_0$ smaller than $\phi_c(0)$ corresponding to the maximum mass, are stable, whereas
the ones with bigger values than $\phi_c(0)$ are unstable.
Thus we expect that the value of $\varphi_0$ corresponding 
to the maximal mass is the boundary between 
stable and unstable equilibrium configurations.

Of course, a complete proof still requires an analysis of the eingenvalues
of the pulsation equations for given ($D, \Lambda)$ 
in order to establish if they are real and positive 
for central densities smaller than $\phi_c(0)$.
However, it is a tedious procedure to find the characteristic values of
$\chi^2$ for a given value of $\phi_0$ (see \cite{Hawley:2000dt} for a detailed
 discussion of this problem in $D=4,~\Lambda=0$ case).
For our case,
we solved the equations (\ref{p1-f}), (\ref{p2-f}) in a neighbourhood of $\phi_c(0)$,
for $D=4$ and
several negative values of $\Lambda$ (we consider nodeless unperturbed solutions only).
The boundary conditions are as follows: 
for $r \to 0$ we find
\begin{eqnarray}
&&f_1=p_0+p_2r^2+O(r^3),~~g'=q_1r+O(r^2),
\\
\nonumber
&&p_2=\omega \varphi_0q_1+\frac{p_0}{2(D-1)}
\left(-\chi^2 e^{-\nu_0(0)}+\mu^2+3e^{-\nu_0(0)}\omega^2\right).
\end{eqnarray}
We can use the linearity of the equations  (\ref{p1-f}), (\ref{p2-f}) to set $f_1(0)=1$.
Dimensionless quantities are obtained by using the same rescaling discussed in Section 2. 
We then have two parameters left $\chi^2$ and $q_1$ 
and the solution of the problem can be found by using a standard shooting procedure.
The numerical results in this case confirm that $\chi^2>0$ for central values of the 
scalar field  smaller than $\phi_c(0)$ while other configurations are unstable ($\chi^2<0$).
In Figure 5 we present the result of the numerical integration for the perturbation
of the metric function $\lambda$ for three diferent 
configurations with maximal mass ($\chi^2=0$).
The variation of $\chi^2$ with the central value of the scalar field is presented in
Figure 6 for two negative values of the cosmological constant.
We have all reasons to expect to find a similar result for any $\Lambda \leq 0 $ and $D>3$.

\section{THE MASS AND THE BOUNDARY STRESS TENSOR}
An important problem of AdS space concerns the definition of mass and angular momentum of 
AAdS spacetimes.
The generalization of Komar's formula in this case is not 
straightforward and 
requires the further subtraction of a background configuration in order 
to render a finite result. 
This problem 
was addressed for the first time in the eighties,
with different approaches (see for instance 
Ref. \cite{Abbott:1982ff, Henneaux:1985tv}).
 
At spatial infinity, the line element (\ref{metric}) can be written as
\begin{equation}
\label{mass1}
ds^2=ds_0^2+h_{\mu \nu}dx^{\mu \nu},
\end{equation}
where $h_{\mu \nu}$ are deviations from the background AdS metric $ds_0^2$.
Similar to the asymptotically flat case, one expects 
the values of mass and other conserved quantities to be encoded 
in the functions $h_{\mu \nu}$.

Using the Hamiltonian formalism, Henneaux and Teitelboim \cite{Henneaux:1985tv}
have computed the mass of an AAdS spacetime in the following way.
They showed that the Hamiltonian must be supplemented by surface terms
in order to be consistent with the equations of motion.
These surface terms yield conserved charges associated with the Killing vectors
of an AAdS geometry.
The total energy is the charge associated with the Killing vector
 $\partial/\partial t$. 
A similar result has been obtained by Abott and Deser \cite{Abbott:1982ff}, 
in a more general context
(see Ref. \cite{Chrusciel:2001qr} for a discussion of the relation
between the Hamiltonian mass and the Abott-Deser one).
This method requires to choose a reference background with suitable matching conditions.
However, in some interesting cases (such as configurations with NUT charge) this choice is ambiguous.

Another formalism to define conserved charges in AAdS 
spacetimes was proposed in \cite{Ashtekar}
and uses conformal techniques to construct conserved quantities.
This construction makes no reference to an action. 
However, it yields the results obtained by Hamiltonian methods.

A different approach has been proposed recently for locally AdS$_4$ asymptotically geometries
and generalized in \cite{Aros:1999kt} for  asymptotically AdS$_{2n}$ spacetimes  (see also \cite{Mielke:eh}).
It consists 
in adding an Euler (boundary-)term to the gravitational
action \cite{Aros:1999id}. The purpose is to guarantee that the action reaches an extremum 
for solutions which are locally AdS at the boundary. 
The resulting action gives the mass and angular momentum 
as Noether charges associated to the asymptotic Killing 
vectors $\partial/\partial t$ and $\partial/\partial \varphi$.
In this formalism the specification of a reference background is also unnecessary.
The efficiency of this approach has been demonstrated in a broad range of examples, includding 
Kerr-AdS and Reissner-Nordstr\"om-AdS black holes, black strings and Taub-NUT and Taub-Bolt solutions.

As expected, these different methods yield the same total mass for 
the spherically symmetric AAdS boson star solutions considered in this paper
(for computations using Hamiltonian methods we consider the obvious vacuum AdS$_D$ background)
\begin{equation}
\label{madm}
M_{ADM}=\frac{(D-2)\Omega_{D-2}}{8 \pi G_D}M,
\end{equation}
where $\Omega_{D-2}=2 \pi^{(D-1)/2}/\Gamma((D-1)/2)$ is the area of a unit $(D-2)$-dimensional sphere.

A procedure leading (for odd dimensions) to a different
result has been proposed by  Balasubramanian
and Kraus \cite{Balasubramanian:1999re}.
This technique was inspired by AdS/CFT correspondence and consists also in adding suitable counterterms 
$I_{ct}$
to the action in order to ensure the finiteness of the stress
tensor derived by the quasilocal energy definition \cite{Brown:1993br}. 
These counterterms are built up with
curvature invariants of a boundary $\partial \cal{M}$ 
(which is sent to infinity after the integration)
and thus obviously they do not alter the bulk equations of motion.

Given the potential relevance of these boson star solutions in an AdS/CFT context, 
we present here a computation of the boundary stress tensor and of the total mass, by using the
counterterm prescription.

The following counterterms are sufficient to cancel divergences 
in a pure gravity theory for 
$D \leq 6$, with several exceptions (see e. g. \cite{Taylor-Robinson:2000xw}):
\begin{eqnarray}
\label{ct}
I_{\rm ct}=-\frac{1}{8 \pi G} \int_{\partial {\cal M}}d^{D-1}x\sqrt{-h}\Biggl[
\frac{D-2}{l}+\frac{l}{2(D-3)}\cal{R}
\Bigg]\ .
\end{eqnarray}
Here ${\cal R}$ is the Ricci scalar for the boundary metric $h$. 
Our conventions with respect to indices will be: $\{A,B,\ldots \}$ indicate 
the intrinsic coordinates of the boundary and $\{a,b,\ldots \}$ indicate the angular coordinates.
In this section we will define also $l^2=- (D-2)(D-1)/(2\Lambda)$. 

Using these counterterms one can
construct a divergence--free stress tensor from the total action
$I{=}I_{\rm bulk}{+}I_{\rm surf}{+}I_{\rm ct}$, by defining (see {\it
e.g.}  Ref.\cite{Brown:1992br})
\begin{eqnarray}
\label{s1}
T_{AB}&=& \frac{2}{\sqrt{-h}} \frac{\delta I}{ \delta h^{AB}}
=\frac{1}{8\pi G}(K_{AB}-Kh_{AB}-\frac{D-2}{l}h_{AB}+\frac{l}{D-3}E_{AB}),
\end{eqnarray}
where $E_{AB}$ is the Einstein tensor of the intrinsic metric $h_{AB}$.

We can use this result to assign a mass to an AAdS geometry
by writing the boundary metric in an ADM form
\begin{eqnarray}
\label{b-AdS}
h_{AB}dx^{A} dx^{B}=-N_{\Sigma}^2dt^2
+\sigma_{ab}(dx^a+N_{\sigma}^a dt) (dx^b+N_{\sigma}^b dt)
\end{eqnarray}
and the definition
\begin{eqnarray}
\label{mass}
\bar{M}=\int_{\partial \Sigma}d^{D-1}x\sqrt{\sigma}N_{\Sigma}\epsilon.
\end{eqnarray}
Here $\epsilon=u^{\mu}u^{\nu}T_{\mu \nu}$ is the proper energy density while $u^{\mu}$ 
is a timelike unit normal to $\Sigma$.

The metric restricted to the boundary $h_{AB}$ diverges due to an infinite
conformal factor $r^2/l^2$. The background metric upon which the dual field
theory resides is
\begin{eqnarray}
\gamma_{AB}=\lim_{r \rightarrow \infty} \frac{l^2}{r^2}h_{AB}.
\end{eqnarray}
For the asymptotically $AdS_D$ solutions considered here, the $(D-1$) 
dimensional boundary is the Einstein universe,
with the line element
\begin{eqnarray}
\label{b-metric}
\gamma_{AB}dx^A dx^B=-dt^2+l^2d\Omega^2_{D-2}.
\end{eqnarray}

If there are matter fields on $\cal{M}$, additional counterterms may be needed to regulate the action.
However, we find that for a boson star spacetime in $D=3,4,5$ dimensions, 
the prescription  (\ref{ct}) removes all divergences 
(this is a consequence of the asymptotic behavior of the scalar field (\ref{asymptotics})).
The results we find for the boundary stress tensor of a boson star at large $r$ 
by using the asymptotic expressions (\ref{asymptotics}) 
are 
\begin{eqnarray}
\label{BD3}
T_{ab}^{(3)}&=&l(M-\frac{1}{2})\delta_{ab}+\dots,
\\
T_{tt}^{(3)}&=&\frac{1}{l}(M-\frac{1}{2})+\dots,
\end{eqnarray}
in three dimensions and 
\begin{eqnarray}
\label{BD4}
T_{ab}^{(4)}&=&\frac{lM}{r}\omega_{ab}+\dots
\\
T_{tt}^{(4)}&=&\frac{2M}{lr}+\dots,
\end{eqnarray}
for $D=4$. 
The boundary stress tensor for the five-dimensional case is
\begin{eqnarray}
\label{BD5}
T_{ab}^{(5)}&=&(M+\frac{l^2}{8})\frac{l}{r^2}\omega_{ab}+\dots
\\
T_{tt}^{(5)}&=&\frac{3M}{lr^2}+\frac{3l}{8r^2}+\dots,
\end{eqnarray}
By using the relation (\ref{mass}), we can find the mass of boson star solutions, according to 
Balasubramanian and Kraus.
The mass of a four-dimensional boson star is $\bar{M}=M$, predicted also by (\ref{madm}).
The result for $D=3$ is $\bar{M}=M-1/2$, while in five dimensions we find 
$\bar{M}=M_{ADM}+ 3\pi l^2/32$.
However, in $D=5$, the standard value is $M_{ADM}$. The additional constant $3\pi l^2/32$
is the mass of pure global $AdS_5$ and is usually interpreted as the energy dual to the Casimir energy 
of the CFT defined on a four dimensional Einstein universe \cite{Balasubramanian:1999re}.

For $D=3$, we find that the mass of solutions computed in this way, 
is always larger than $-1$, with the extreme value 
$\bar{M}=-1$ corresponding to the global $AdS_3$ space. Contrary to what happens
for BTZ black holes, we do not find a mass gap between $\bar{M}=-1$ and $\bar{M}=0$.
To compute the mass for this type of configuration, usually one uses a matching procedure 
on a surface separating the regions where 
the internal and external geometries are defined (see $e.g.$ \cite{Sakamoto:1998hq}). 
The external geometry is taken to be the BTZ black hole which gives a value for the mass of the solutions.
The counterterm  method gives, however, a rigurous prescription to compute
the total mass of a 3D star, without the rather unnatural matching procedure.

In light of the AdS/CFT correspondence, Balasubramanian and Kraus have interpreted 
Eq. (\ref{s1}) as $<\tau^{AB}>=\frac{2}{\sqrt{-\gamma}} \frac{\delta S_{eff}}{\delta \gamma_{AB}}$,
where $<\tau^{AB}>$ is the expectation value of the CFT stress tensor.
Then, the divergences which appear are simply the standard ultraviolet divergences
of a quantum field theory and we can cancel them by adding local counterterms
to the action.
Corresponding to the boundary metric (\ref{b-metric}), the stress-energy tensor $\tau_{AB}$
for the dual theory can be calculated using the following relation \cite{Myers:1999qn}
\begin{eqnarray}
\label{r1}
\sqrt{-\gamma}\gamma^{AB}<\tau_{BC}>=
\lim_{r \rightarrow \infty} \sqrt{-h} h^{AB}T_{BC}.
\end{eqnarray}

However, further progress in this direction seems difficult, since
 we do not know the underlying boundary CFT for the action (\ref{action}).
In particular, we do not know what the gravitating complex scalar field corresponds 
to in CFT language.

\section{CONCLUDING REMARKS}

Many particle theories predict that weakly interacting bosonic particles
are abundant in the universe, and that dark matter can be bosonic in nature.

In this paper we investigated the basic properties of boson star solutions 
in the presence of a negative
cosmological constant.
Both analytical and numerical arguments have been presented for the existence of nontrivial
solutions. A general virial relation has been found and particular cases have been discussed.

The results we find for $D\geq4$ are broadly similar to those valid for the 
known asymptotically flat, four dimensional case and there are only some small differences.
Nontrivial solutions are again found for every value of the
central scalar field and discrete values of frequencies while the scalar field vanishes at infinity.
However, a nonzero $\Lambda$  term in the action implies a complicated power decay of the 
fields at infinity, rather than exponentially as expected.
Also the parameters of the solutions in the asymptotically flat case 
(mass, particle number, effective radius)
are no 
longer valid and take new values for every choice of $\Lambda$. 
The mass of the star decreases with the cosmological constant strength
$|\Lambda|$.
Therefore the existence of a negative cosmological constant implies a decrease in the 
maximal mass of the star.
In the sense of the no-scalar-hair theorem we have shown that, for this system, 
the only spherically symmetric 
black hole solution in $D-$dimensions is the Schwarzschild-AdS solution.

These results are not so surprising, since the different asymptotic structure
of spacetime obtained for a $\Lambda \neq 0$ does not manifest here directly.
The existence of the term $V(\Phi)\sqrt{-g}$ in the  action (\ref{action})
still implies $\Phi \to 0$ asymptotically (since $g_{rr}g_{tt} \to -1$),
independently on the presence of $\Lambda$.

A similar mechanism will act also for a Higgs scalar interacting with 
a nonabelian field in an AAdS geometry \cite{Lugo:1999fm,VanderBij:2001ah}.
Here we consider only the better known case of a SU(2) field in four dimensions.
In this case, a spherically symmetric purely magnetic
YM field can be written as (see e.g. \cite{Bjoraker:2000qd})
\begin{eqnarray}
\label{w}
A=\frac{1}{2}\Big [\omega(r)\tau_1 d \theta+\left(\cot \theta \tau_3+\omega(r) \tau_2\right) 
\sin \theta d \varphi \Big]
\end{eqnarray}
(where $\omega$ is the YM potential and $\tau_i$ are the usual Pauli matrices),
which implies an YM effective lagrangian 
$L_{YM}=1/2e^{-\delta}(F \omega^{\prime 2}+V(\omega))$, with $V(\omega)=(\omega^2-1)^2/2r^2$.
The scalar field effective lagrangian is 
$L_{\phi}=r^2 e^{-\delta}(F \phi^{\prime 2}+V(\phi))$, with $V(\phi)$ the standard Higgs potential.
The lagrangian of the system contains also an interacting term
$L_i=e^{-\delta}T(\omega)\phi^2$, where $T(\omega)=(\omega+1)^2$ for a doublet Higgs field 
\cite{VanderBij:2001ah}
and $T(\omega)=\omega^2$ for a  Higgs field in the adjoint representation 
\cite{Lugo:1999fm}.
Again the finite energy requiement forces  $V(\phi) \to 0$ as $r \to \infty$,
which implies that the Higgs field 
approaches asymptotically the vacuum expectation value.
Moreover, the
interacting term $T(\omega)\phi^2$ implies 
that the allowed values of the
YM potential at infinity are similar to the $\Lambda=0$ case. 
In particular, the solutions are found again for a discrete set of the parameters that specify the initial
conditions at the origin.
Without a scalar field, the YM behavior is very different.
The resulting YM equation reads  
$(e^{-\delta}(F \omega^{\prime}))'= e^{-\delta}\omega (\omega^2-1)/r^2$
and the YM potential $\omega$ takes arbitrary values at infinity $\omega=\omega_0+w_1/r+O(1/r^2)$.
The solutions are found for continuous sets of shooting parameters and have 
a nonzero magnetic charge  \cite{Bjoraker:2000qd}.

Thus, it seems that when studying a gravitating scalar system, the solutions 
exhibit a generic behavior for $\Lambda \leq 0$. Given the presence of a cosmological event horizon, 
the case $\Lambda>0$ needs a separate analysis.

We noticed, however, that there is a curious change of behaviour for $D=3$, 
with somewhat different solution properties.

We expect also to obtain a very similar qualitative behavior of the solutions
when discussing a number of possible
extensions of this theory, $e.g.$ including a dilaton or a 
selfinteraction term.
\\
\\
{\bf Acknowledgement}
\\
We would like to thank J.J. van der Bij  
for valuable discussions and Robert C. Myers for his critical reading of our paper.
We thank Dave Winters for proof-reading an earlier draft of this paper.

The work of one of the authors (E.R.) was performed in the context of the
Graduiertenkolleg of the Deutsche Forschungsgemeinschaft (DFG):
Nichtlineare Differentialgleichungen: Modellierung,Theorie, Numerik, Visualisierung. 
D.A. would like to thank the Perimeter Institute for Theoretical 
Physics and the University of Waterloo's Department of Physics 
for their kind hospitality during this work.
\\ 


%
%
%
%
\newpage
{\bf Figure Captions}
\\
\\
\\
Figure 1:
\\
The mass-parameter $M$   (dotted line)  and the particle number $N$  
(solid line) are represented as a function of $\phi(0)$ for typical 
boson stars with $\Lambda\leq0$ in three- (figure 1a), four (figure 1b) and five dimensions (figure 1c).
\\
In figures 1-4 the mass parameter $M$ is given in units 
$\frac{(D-2)\pi^{(D-3)/2}}{4\Gamma\big (D-1)/2 \big) }\frac{M_{Pl}^{D-2}}{\mu^{D-3}}$,
while the particle number $N$ is given in units 
$\frac{(D-2)\pi^{(D-3)/2}}{4\Gamma\big (D-1)/2 \big) }(\frac{M_{Pl}}{ \mu})^{D-2}$.
\\
\\
\\
Figure 2:
\\
The mass-parameter $M$  is represented 
as a function of the negative cosmological constant for
different values of the central scalar field $\varphi_0$ for 
boson star solutions in three- (figure 2a), four (figure 2b) and five dimensions (figure 2c).  
\\
\\
\\
Figure 3:
\\
The particle number $N$ is represented 
as a function of the negative cosmological constant for
different values of the central scalar field $\varphi_0$ for 
boson star solutions in three- (figure 3a), four (figure 3b) and five dimensions (figure 3c).  
\\
\\
\\
Figure 4:
\\
The particle number $N$ is represented 
as a function of the effective radius $R$ (in units $1/\mu$) for
different values of the  cosmological constant
in three- (figure 4a), four (figure 4b) and five dimensions (figure 4c).    
\\
\\
\\
Figure 5:
\\
The perturbation of the metric function $\delta \lambda$ is
shown as a function of the radial coordinate for the fundamental mode
and several values of $\Lambda$.
\\
\\
\\
Figure 6:
\\
The squared frequency $\chi^2$ of the perturbations is plotted
as a function of scalar field at the origin 
for $D=4$  asymptotically AdS boson star solutions with 
$\Lambda=-0.05,~-0.5$.
Note that $\chi^2$ is zero when $\Phi(0)\simeq 0.48$ for $\Lambda=-0.05$ and
 $\Phi(0)\simeq 0.652$ for $\Lambda=-0.5$, which corresponds to boson stars with
 maximum possible mass.

\newpage
\setlength{\unitlength}{1cm}

\begin{picture}(16,16)
\centering
\put(-2,0){\epsfig{file=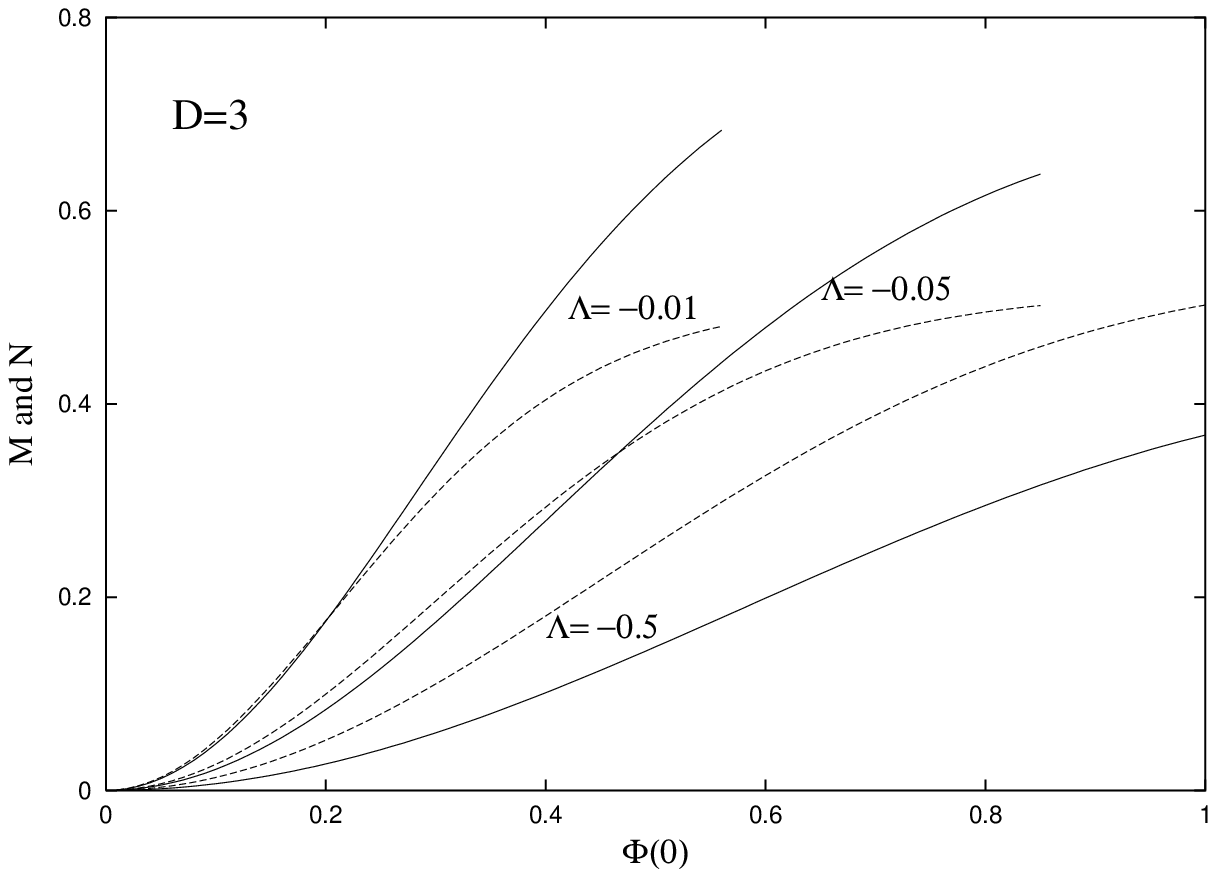,width=18cm}}
\end{picture}
\begin{center}
Figure 1a.
\end{center}

\newpage
\setlength{\unitlength}{1cm}

\begin{picture}(16,16)
\centering
\put(-2,0){\epsfig{file=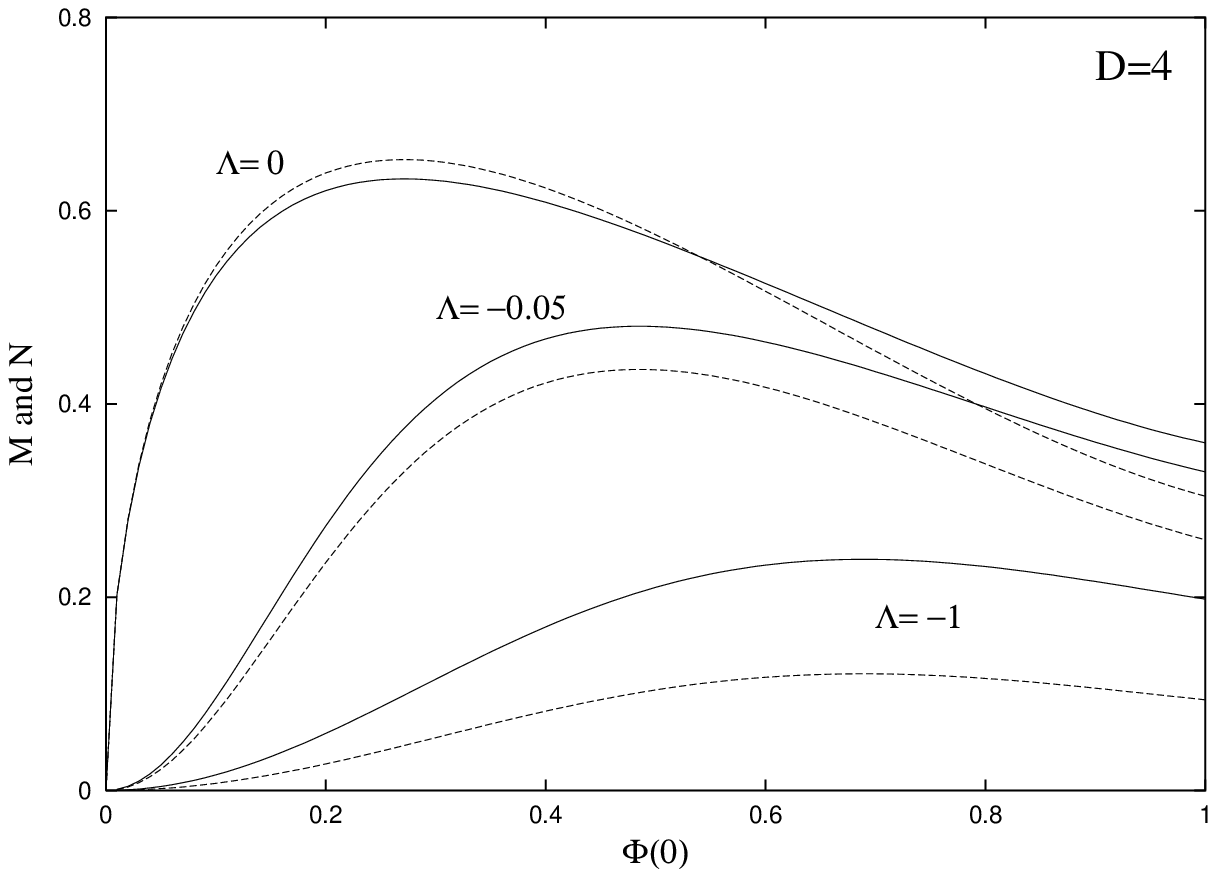,width=18cm}}
\end{picture}
\begin{center}
Figure 1b.
\end{center}

\newpage
\setlength{\unitlength}{1cm}

\begin{picture}(16,16)
\centering
\put(-2,0){\epsfig{file=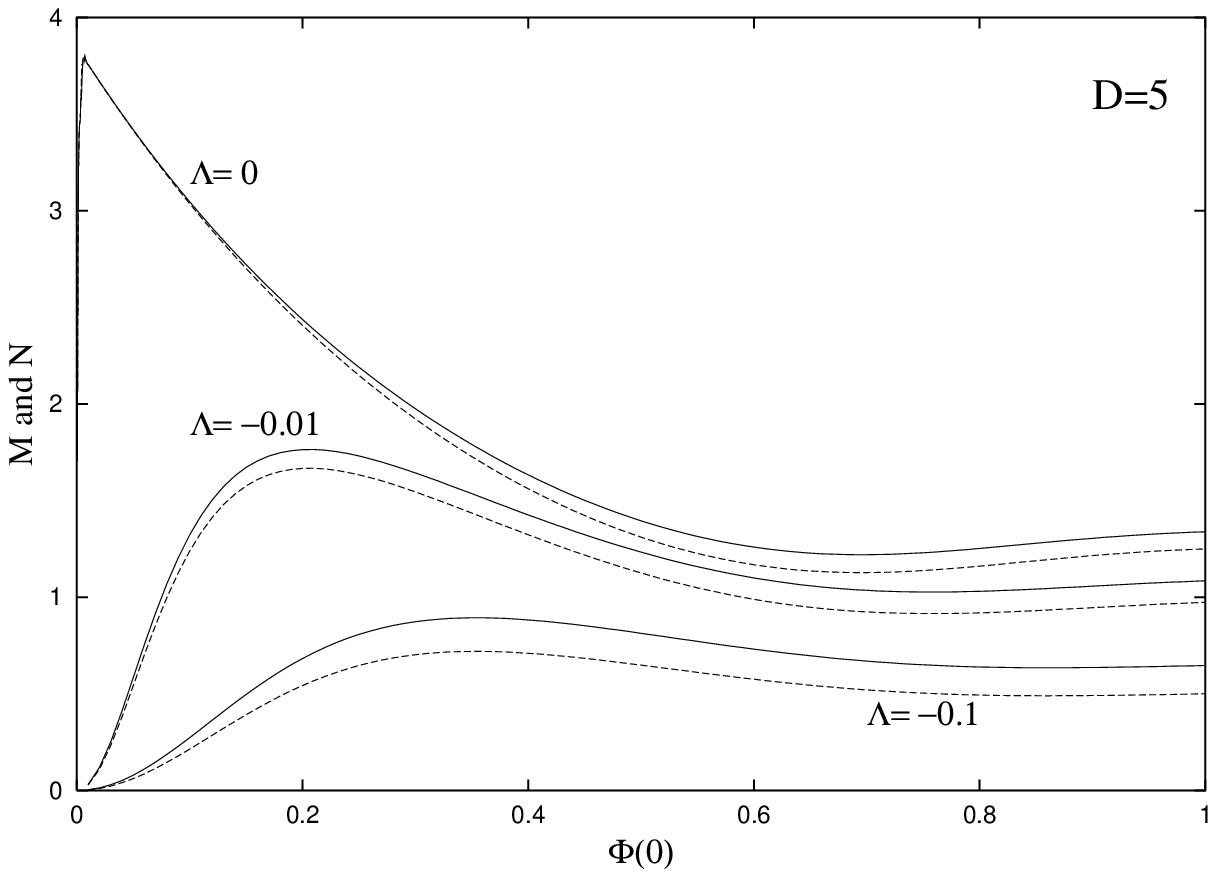,width=18cm}}
\end{picture}
\begin{center}
Figure 1c.
\end{center}

\newpage
\setlength{\unitlength}{1cm}

\begin{picture}(16,16)
\centering
\put(-2,0){\epsfig{file=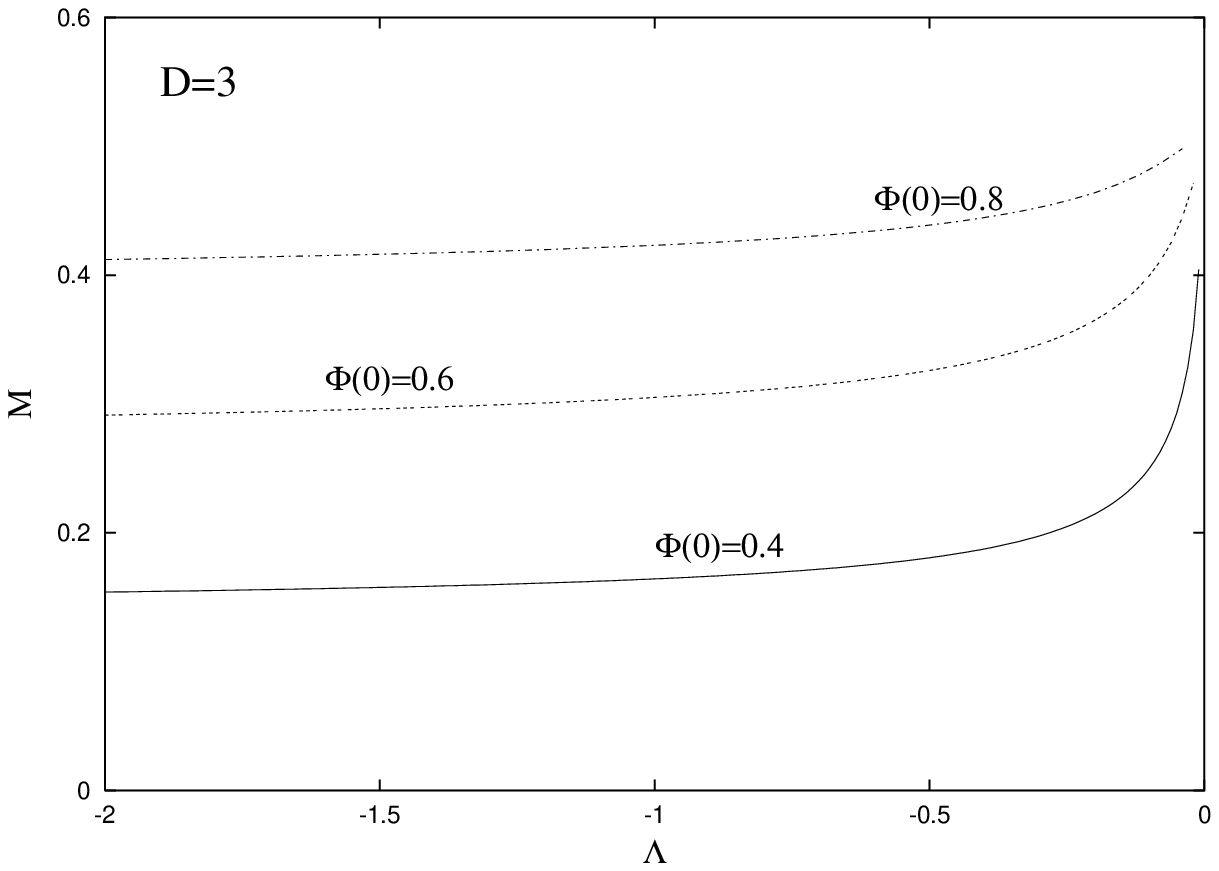,width=18cm}}
\end{picture}
\begin{center}
Figure 2a.
\end{center}

\newpage
\setlength{\unitlength}{1cm}

\begin{picture}(16,16)
\centering
\put(-2,0){\epsfig{file=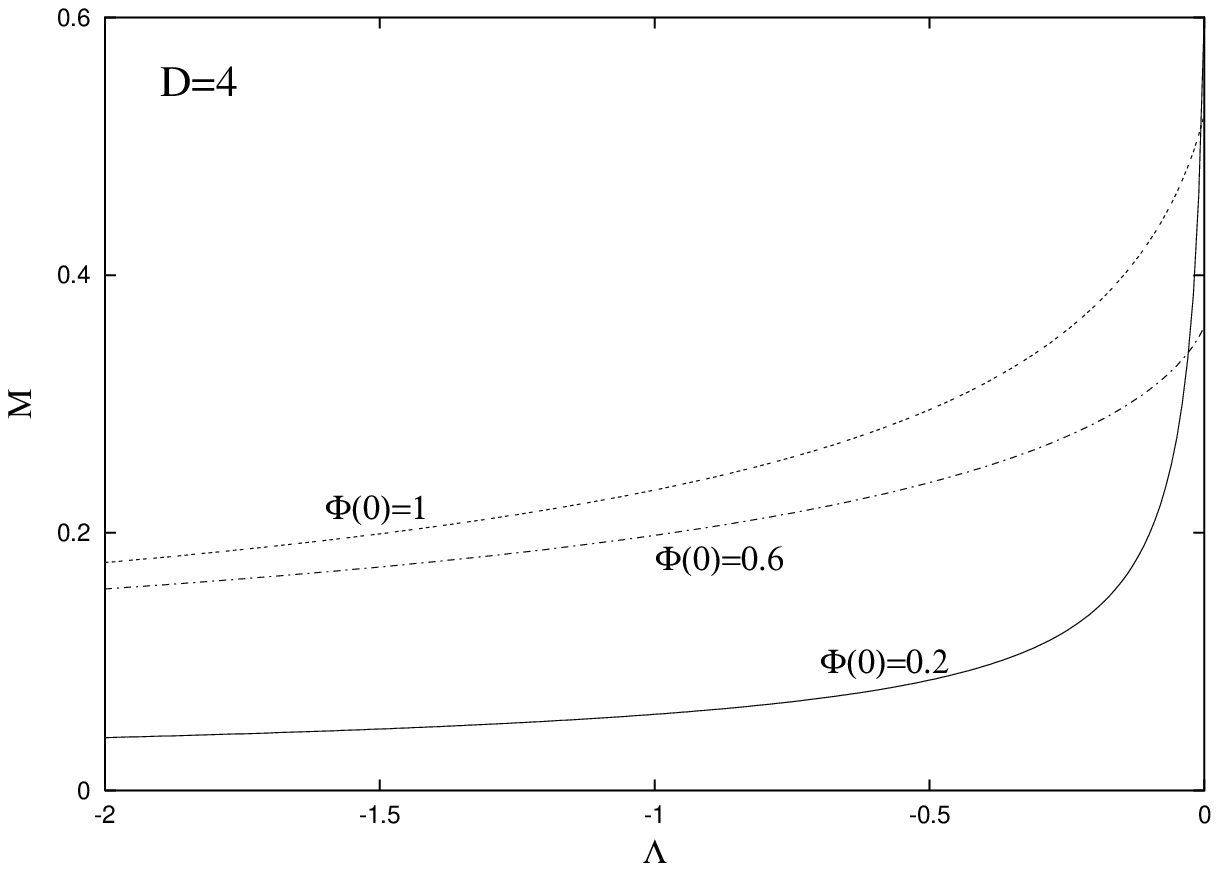,width=18cm}}
\end{picture}
\begin{center}
Figure 2b.
\end{center}

\newpage
\setlength{\unitlength}{1cm}

\begin{picture}(16,16)
\centering
\put(-2,0){\epsfig{file=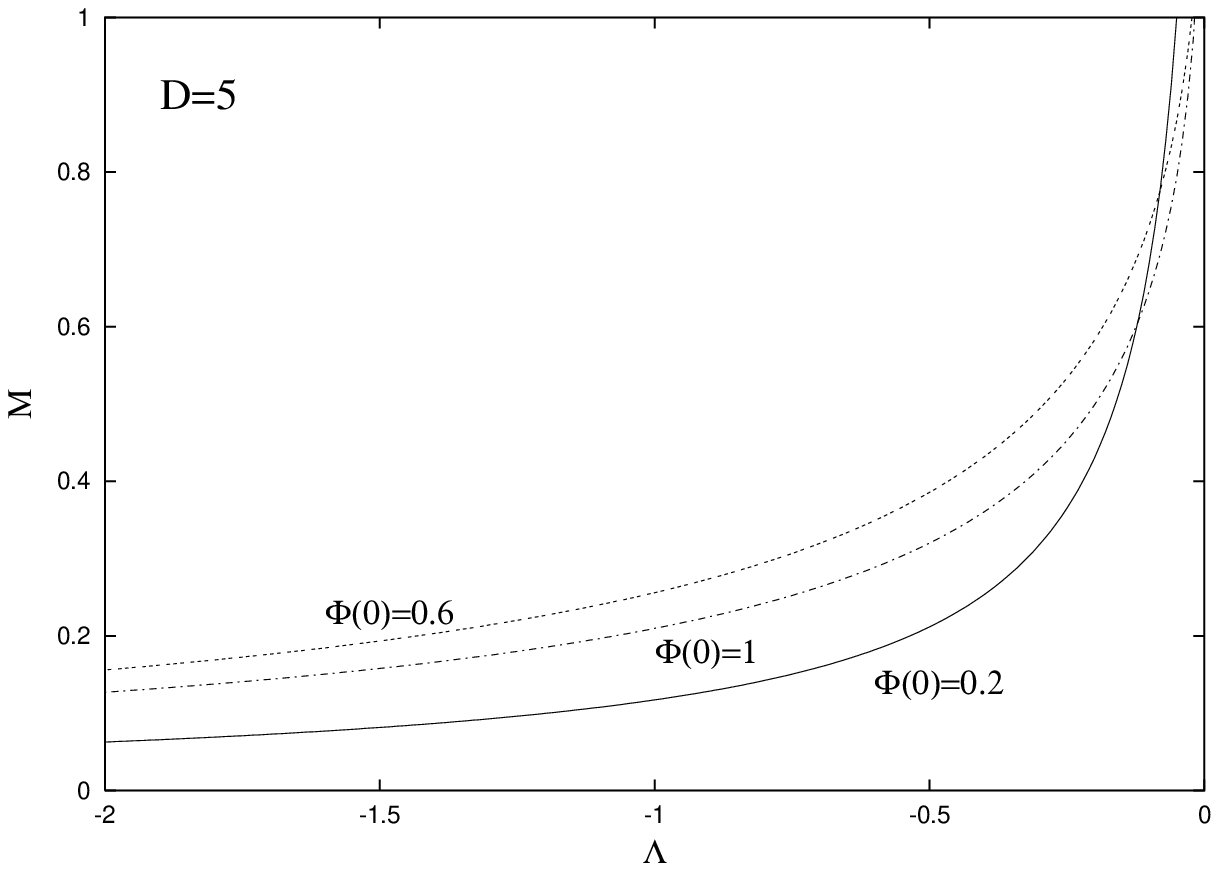,width=18cm}}
\end{picture}
\begin{center}
Figure 2c.
\end{center}

\newpage
\setlength{\unitlength}{1cm}

\begin{picture}(16,16)
\centering
\put(-2,0){\epsfig{file=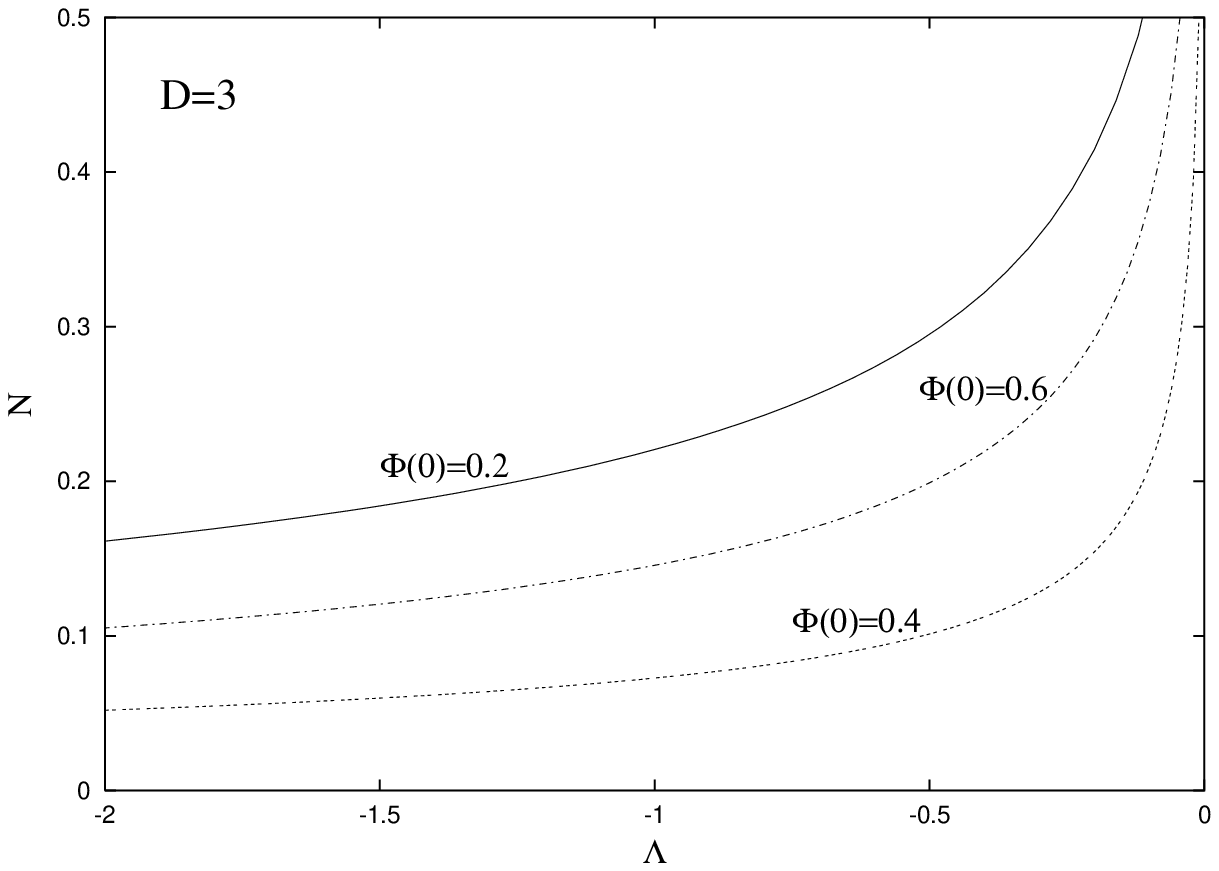,width=18cm}}
\end{picture}
\begin{center}
Figure 3a.
\end{center}

\newpage
\setlength{\unitlength}{1cm}

\begin{picture}(16,16)
\centering
\put(-2,0){\epsfig{file=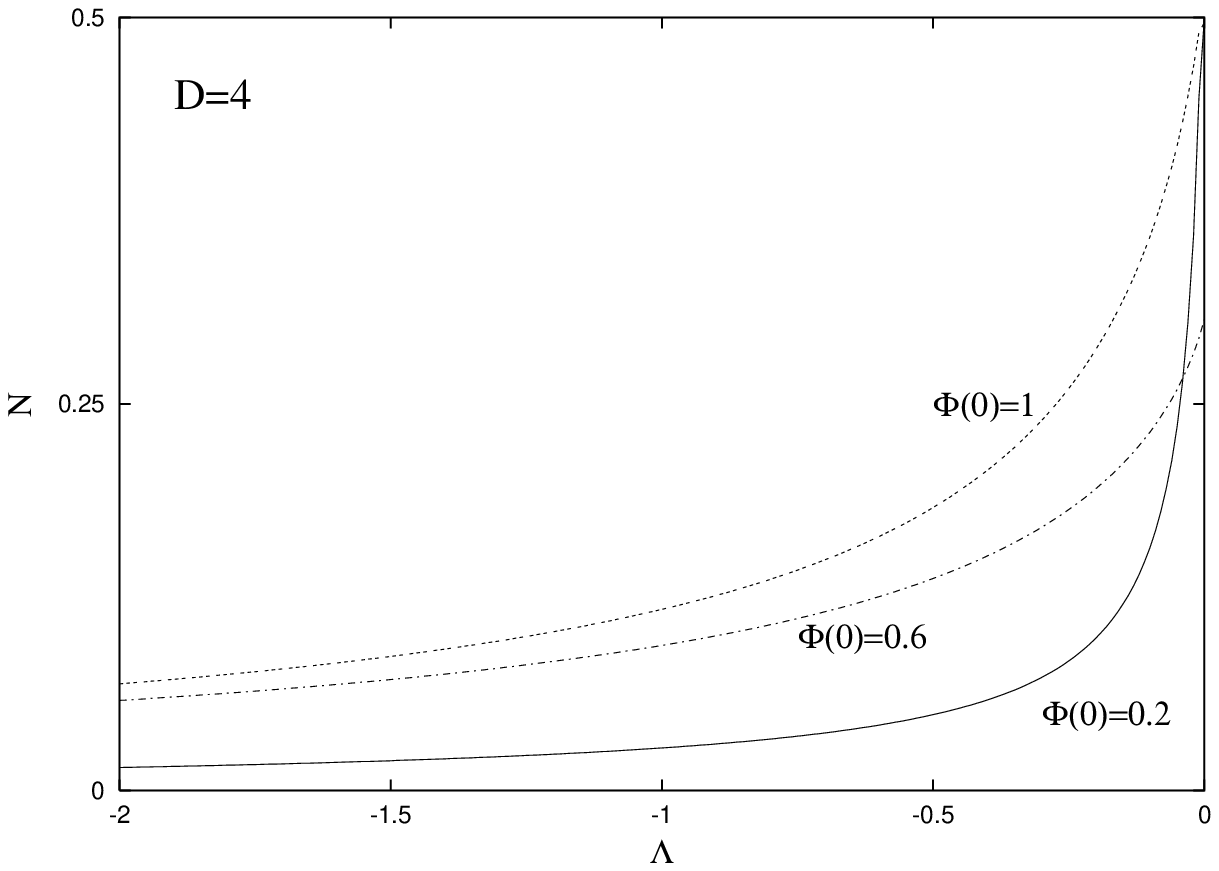,width=18cm}}
\end{picture}
\begin{center}
Figure 3b.
\end{center}

\newpage
\setlength{\unitlength}{1cm}

\begin{picture}(16,16)
\centering
\put(-2,0){\epsfig{file=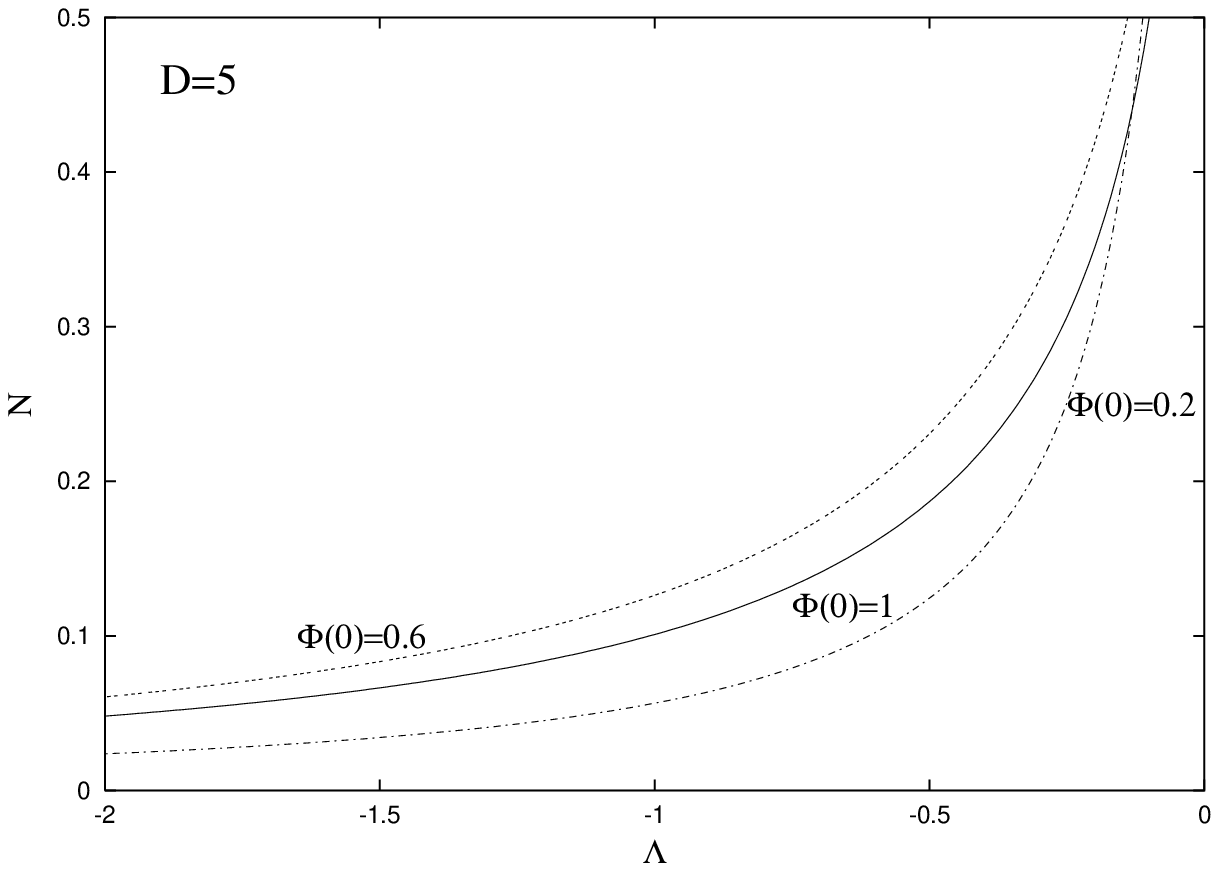,width=18cm}}
\end{picture}
\begin{center}
Figure 3c.
\end{center}

\newpage
\setlength{\unitlength}{1cm}

\begin{picture}(16,16)
\centering
\put(-2,0){\epsfig{file=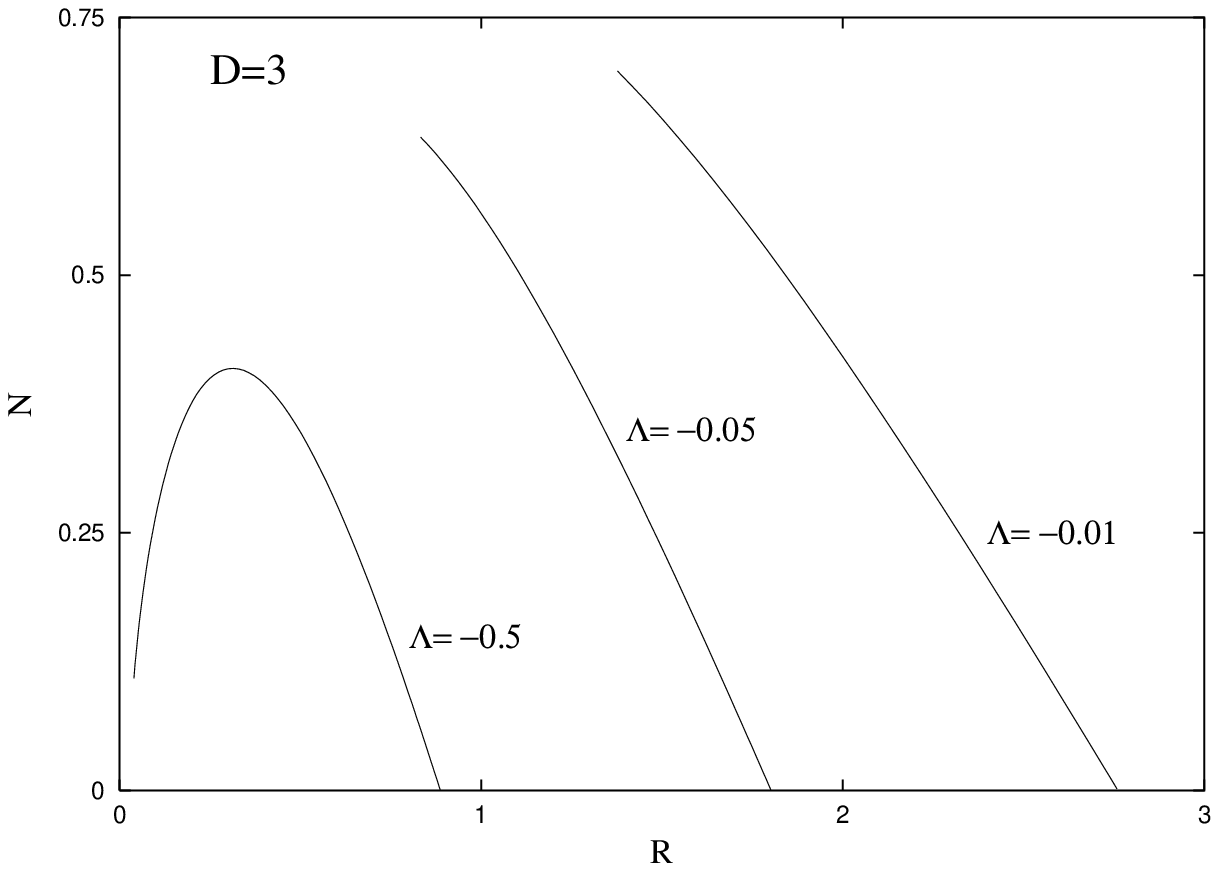,width=18cm}}
\end{picture}
\begin{center}
Figure 4a.
\end{center}

\newpage
\setlength{\unitlength}{1cm}

\begin{picture}(16,16)
\centering
\put(-2,0){\epsfig{file=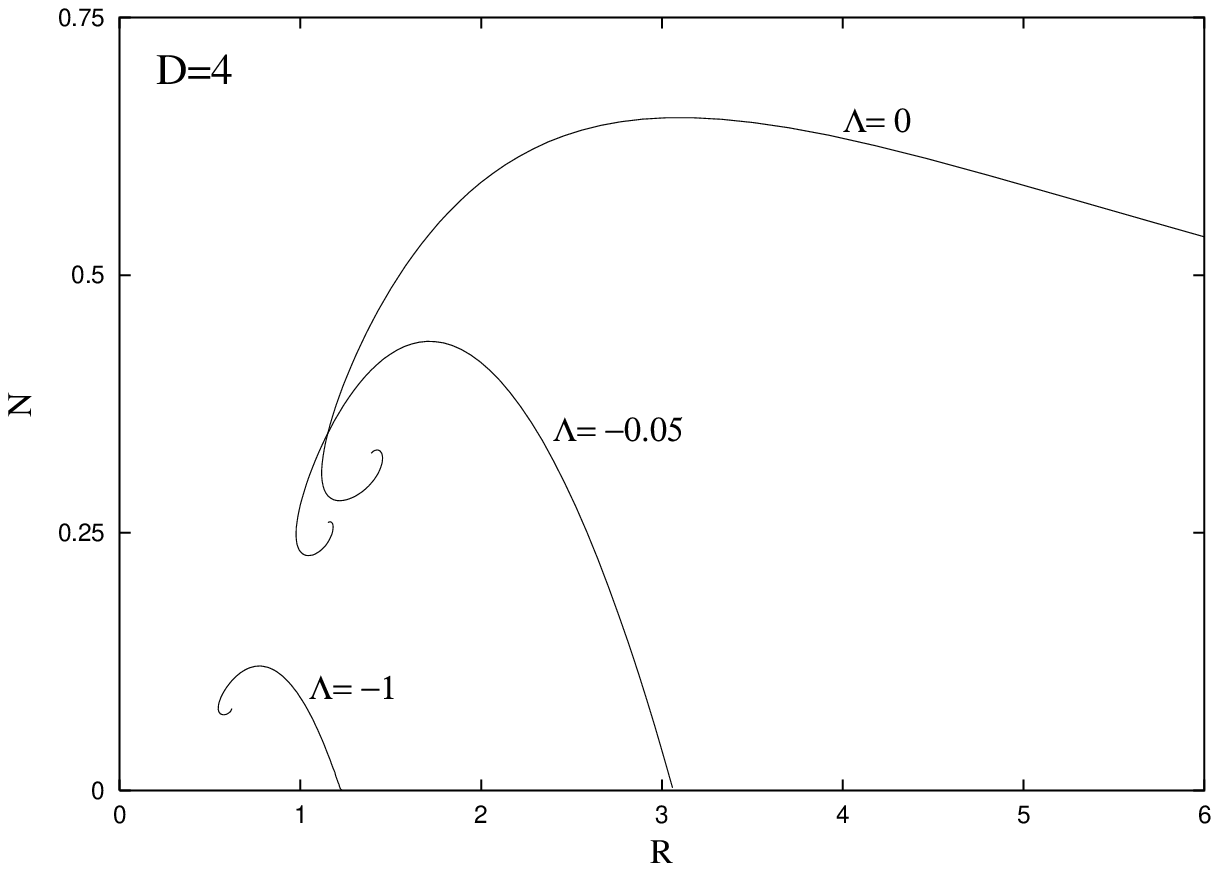,width=18cm}}
\end{picture}
\begin{center}
Figure 4b.
\end{center}

\newpage
\setlength{\unitlength}{1cm}

\begin{picture}(16,16)
\centering
\put(-2,0){\epsfig{file=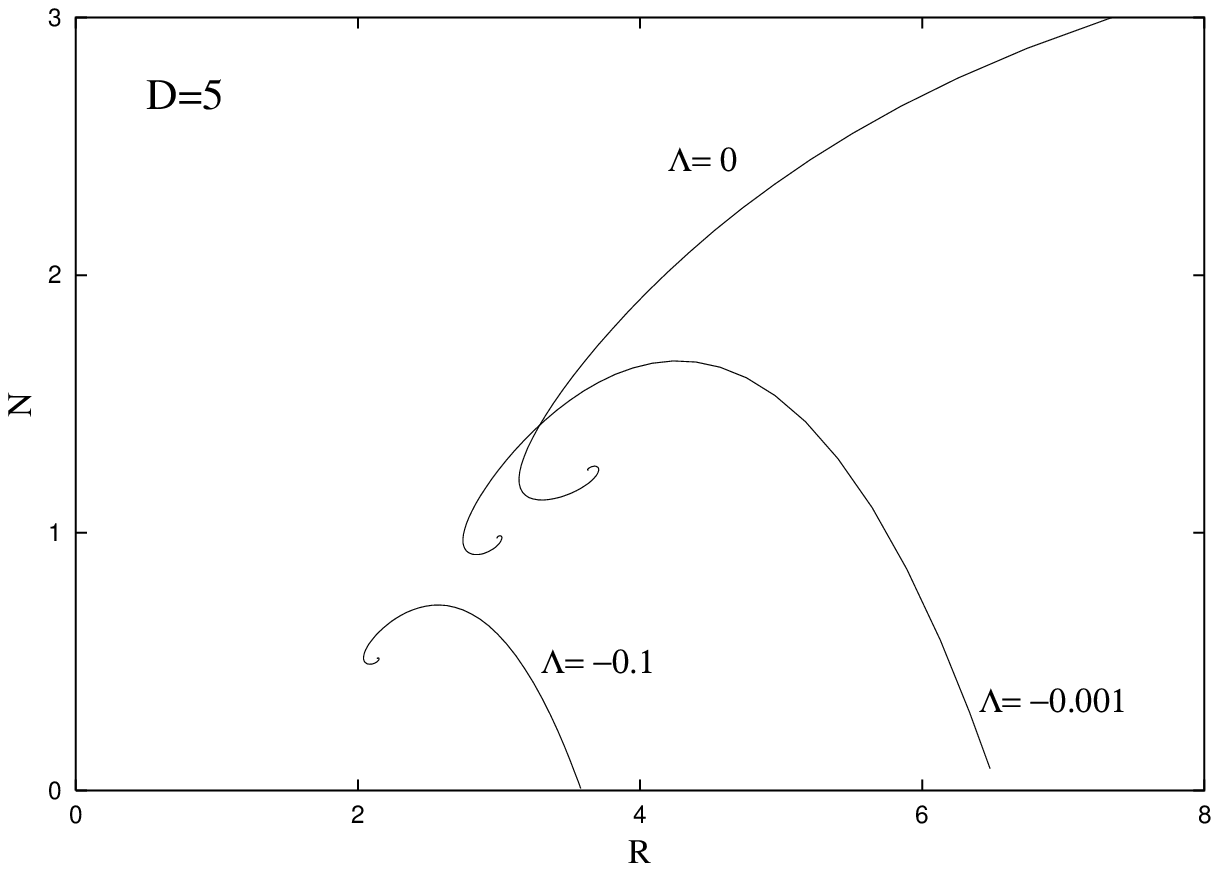,width=18cm}}
\end{picture}
\begin{center}
Figure 4c.
\end{center}

\newpage
\setlength{\unitlength}{1cm}

\begin{picture}(16,16)
\centering
\put(-2,0){\epsfig{file=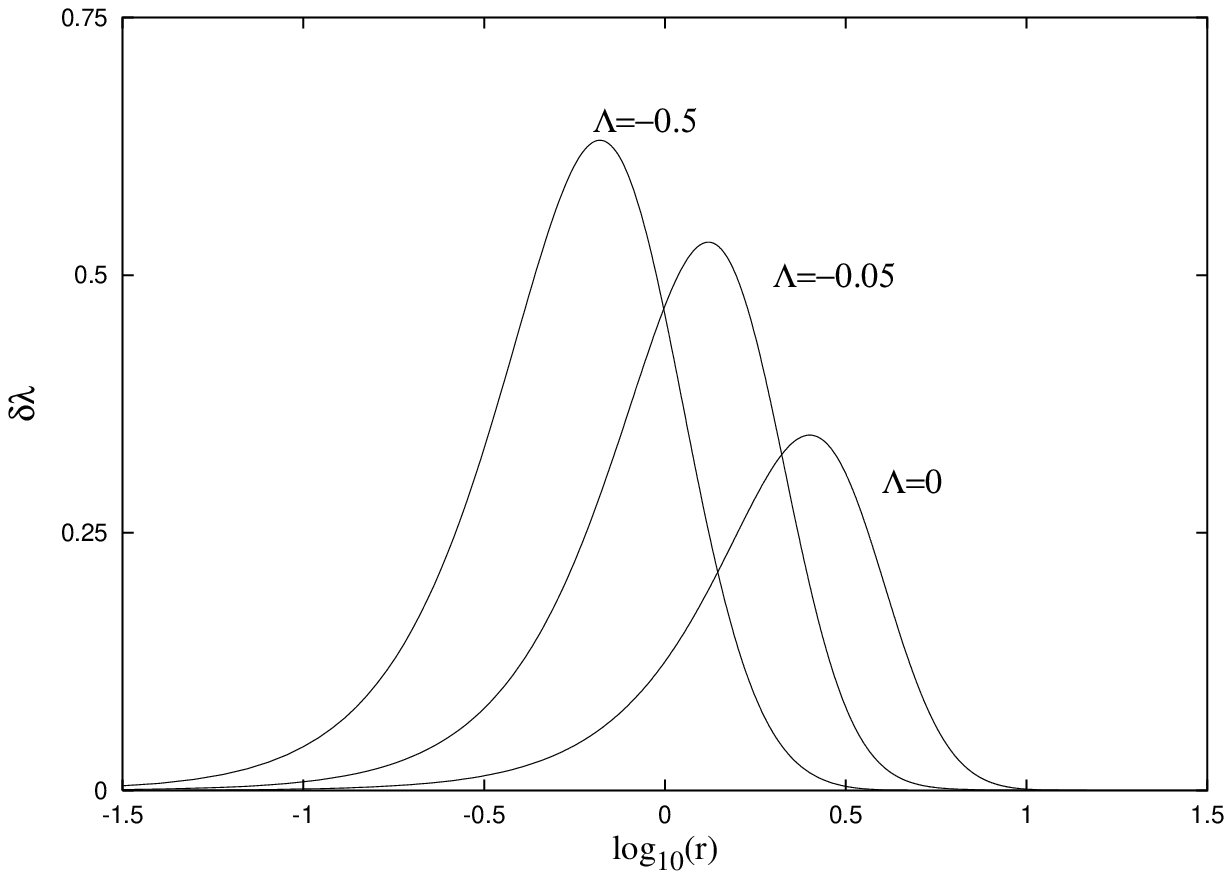,width=18cm}}
\end{picture}
\begin{center}
Figure 5.
\end{center}

\newpage
\setlength{\unitlength}{1cm}

\begin{picture}(16,16)
\centering
\put(-2,0){\epsfig{file=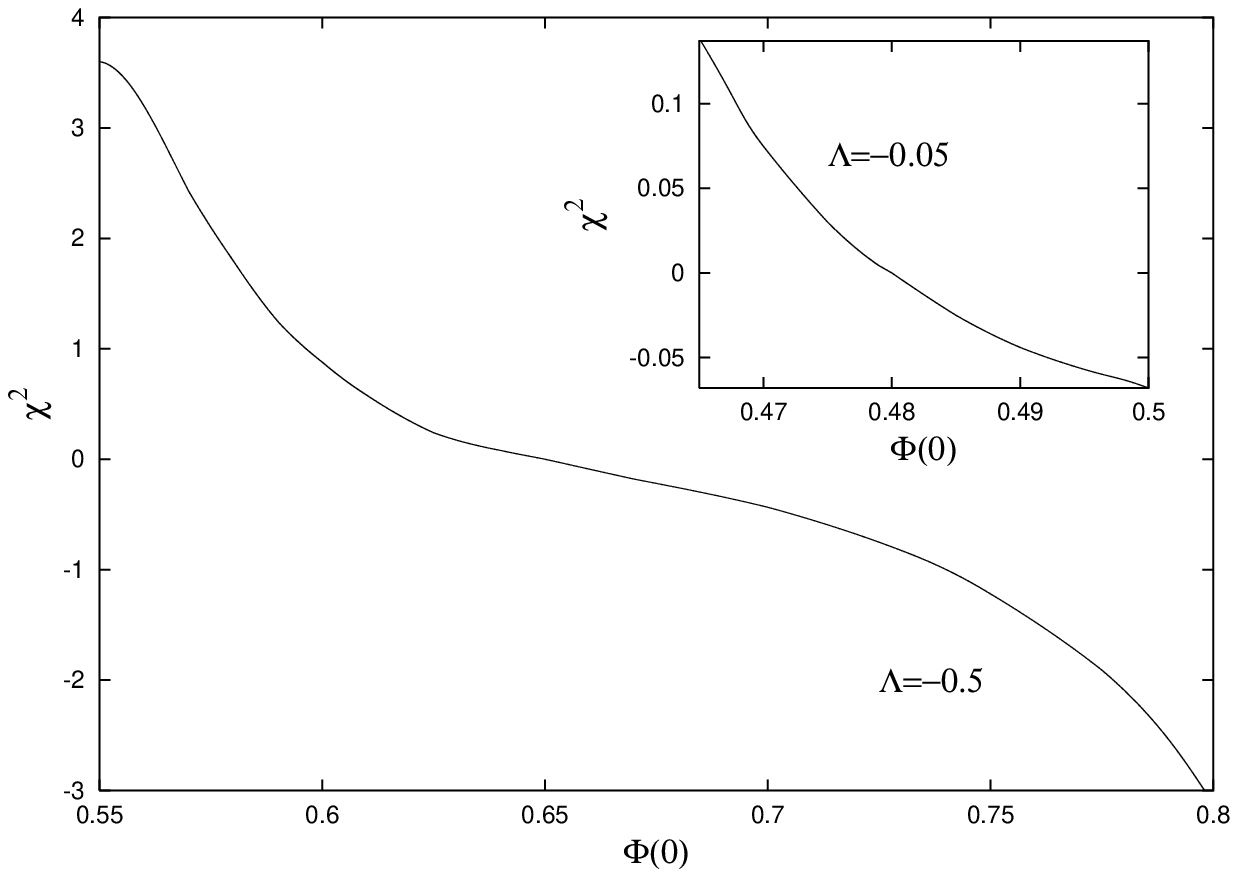,width=18cm}}
\end{picture}
\begin{center}
Figure 6.
\end{center}

\end{document}